\documentclass[twocolumn]{aastex6}
\usepackage{natbib}
\usepackage{enumerate}
\usepackage{import}
\usepackage{xspace}
\usepackage{color}
\usepackage{gensymb}
\usepackage{datetime}
\usepackage[titletoc,title]{appendix}
\usepackage{setspace}
\usepackage{amsmath}

\newcommand{\msinilogP}{$3.28^{+0.16}_{-0.15}$}
\newcommand{\ecclogP}{$0.87\pm0.05$}
\newcommand{\msini}{$3.23^{+0.15}_{-0.14}$}
\newcommand{\aop}{$-0.35\pm0.03$}
\newcommand{\transitprob}{$0.00185\pm0.00010$}
\newcommand{\period}{$74^{+43}_{-22}$}
\newcommand{\gammaK}{$-52.6^{+1.3}_{-1.5}$}
\newcommand{\gammaJ}{$-52.4^{+2.0}_{-2.1}$}
\newcommand{\gammaM}{$-19.2^{+1.9}_{-2.1}$}
\newcommand{\gammaA}{$-47.2^{+2.0}_{-2.2}$}
\newcommand{\jitK}{$3.4^{+0.8}_{-0.6}$}
\newcommand{\jitJ}{$3.3\pm0.4$}
\newcommand{\jitM}{$5.8^{+0.6}_{-0.5}$}
\newcommand{\jitA}{$3.7^{+0.5}_{-0.4}$}
\newcommand{\logper}{$10.21^{+0.46}_{-0.35}$}
\newcommand{\tconj}{$18965^{+44}_{-40}$}
\newcommand{\sesinw}{$-0.32\pm0.03$}
\newcommand{\secosw}{$0.86\pm0.02$}
\newcommand{\logK}{$3.64\pm0.01$}

\newcommand{\semiamp}{$38.25^{+0.58}_{-0.55}$}
\newcommand{\ecc}{$0.84\pm0.04$}
\newcommand{\tp}{$18121\pm12$}
\newcommand{\sma}{$18^{+6}_{-4}$}
\newcommand{\closestapproach}{$2.88^{+0.09}_{-0.08}$}
\newcommand{\teqperi}{$171.0^{+5.2}_{-5.1}$}
\newcommand{\teqapo}{$50.2^{+7.0}_{-7.6}$}
\newcommand{\phaseang}{$137^{+10}_{-19}$}
\newcommand{\gammadot}{$0.12^{+0.22}_{-0.19}$}
\newcommand{\gammadotper}{$71.87^{+52.26}_{-27.07}$}
\newcommand{\gammadotecc}{$0.92\pm0.03$}

\newcommand{\nwalkers}{50}
\newcommand{\nsteps}{1552}
\newcommand{\nensembles}{8}

\newcommand{\maxGR}{1.001}

\newcommand{\nobshires}{78}
\newcommand{\nobsmcd}{175}
\newcommand{\nobsapf}{104}

\newcommand{\mapperiod}{72.85}
\newcommand{\mapecc}{0.84}
\newcommand{\maplogper}{10.2}
\newcommand{\maptc}{18964}
\newcommand{\mapsecosw}{0.86}
\newcommand{\mapsesinw}{-0.32}
\newcommand{\maplogK}{3.64}
\newcommand{\mapjitK}{3.09}
\newcommand{\mapjitJ}{3.16}
\newcommand{\mapjitM}{5.67}
\newcommand{\mapjitA}{3.58}
\newcommand{\mapgammaA}{-47.2}
\newcommand{\mapgammaJ}{-52.4}
\newcommand{\mapgammaM}{-19}
\newcommand{\mapgammaK}{-52.5}
\newcommand{\mapK}{38.21}
\newcommand{\mapaop}{-0.35}
\newcommand{\mapmsini}{3.24}
\newcommand{\mapTp}{18120.0}
\newcommand{\mapa}{18.0}
\newcommand{\mapclosestapproach}{2.89}
\newcommand{\mapteqperi}{170.94}
\newcommand{\mapteqapo}{50.58}
\newcommand{\deltabic}{5.4}
\newcommand{\deltaaic}{1.7}

\newcommand{\msun}{\ensuremath{M_\odot}}
\newcommand{\mjup}{\ensuremath{M_J}}
\newcommand{\ms}{\ensuremath{\mathrm{m\,s^{-1}}}}
\shortauthors{Blunt et al.}
\shorttitle{HR 5183 b}
\submitted{Accepted to AJ}

\begin{document}
\pagenumbering{arabic}

\title{Radial Velocity Discovery of an Eccentric Jovian World Orbiting at 18 au}

\author{ 
    Sarah Blunt\altaffilmark{1,2,3},
    Michael Endl\altaffilmark{4},
    Lauren M. Weiss\altaffilmark{5,6},
    William D. Cochran\altaffilmark{4}, 
    Andrew W.\ Howard\altaffilmark{1},
    Phillip J.\ MacQueen\altaffilmark{4},%
    Benjamin J.\ Fulton\altaffilmark{1, 18},
    Gregory W. Henry\altaffilmark{17},
    Marshall C. Johnson\altaffilmark{15,4}
    Molly R. Kosiarek\altaffilmark{3,7},
    Kellen D. Lawson\altaffilmark{10},
    Bruce Macintosh\altaffilmark{11},
    Sean M. Mills\altaffilmark{1},
    Eric L. Nielsen\altaffilmark{11},
    Erik A. Petigura\altaffilmark{1},
    Glenn Schneider\altaffilmark{9},
    Andrew Vanderburg\altaffilmark{19, 20},
    John P. Wisniewski\altaffilmark{10},
    Robert A.\ Wittenmyer\altaffilmark{16,4}, 
    Erik Brugamyer\altaffilmark{4}, % 
    Caroline Caldwell\altaffilmark{4}, 
    Anita L. Cochran\altaffilmark{4}, 
    Artie P. Hatzes\altaffilmark{13},
    Lea A. Hirsch\altaffilmark{11},
    Howard Isaacson\altaffilmark{8,21},
    Paul Robertson\altaffilmark{14,4},
    Arpita Roy\altaffilmark{1},
    Zili Shen\altaffilmark{4}
}

\altaffiltext{1}{Department of Astronomy, California Institute of Technology, Pasadena, CA, USA}
\altaffiltext{2}{Center for Astrophysics $|$ Harvard \& Smithsonian, Cambridge, MA, USA}
\altaffiltext{3}{NSF Graduate Research Fellow}
\altaffiltext{4}{McDonald Observatory and Department of Astronomy, The University of Texas at Austin, Austin, TX, USA}
\altaffiltext{5}{Institute for Astronomy, University of Hawai`i, Honolulu, HI, USA}
\altaffiltext{6}{Beatrice Watson Parrent Fellow}
\altaffiltext{7}{Department of Astronomy \& Astrophysics, University of California, Santa Cruz, CA, USA}
\altaffiltext{8}{Astronomy Department, University of California, Berkeley, CA, USA}
\altaffiltext{9}{Steward Observatory, The University of Arizona, Tucson, AZ, USA}
\altaffiltext{10}{Homer L. Dodge Department of Physics and Astronomy, University of Oklahoma, Norman, OK, USA}
\altaffiltext{11}{Kavli Institute for Particle Astrophysics and Cosmology, Stanford University, Stanford, CA, USA}
\altaffiltext{13}{Th\"uringer Landessternwarte, Tautenburg, Germany, EU}
\altaffiltext{14}{University of California, Irvine, CA, USA}
\altaffiltext{15}{Department of Astronomy, The Ohio State University, Columbus, OH, USA} 
\altaffiltext{16}{Centre for Astrophysics, University of Southern Queensland, Toowoomba, QLD, Australia}
\altaffiltext{17}{Center of Excellence in Information Systems, Tennessee State University, Nashville, TN, USA}
\altaffiltext{18}{IPAC-NASA Exoplanet Science Institute, Pasadena, CA, USA} 
\altaffiltext{19}{NASA Sagan Fellow} 
\altaffiltext{20}{Department of Astronomy, Unversity of Texas at Austin, Austin, TX, USA}
\altaffiltext{21}{University of Southern Queensland, Toowoomba, QLD 4350, Australia}

\keywords{planets and satellites: detection - planets and satellites: fundamental parameters - stars: individual (HR 5183)}

\begin{abstract}
Based on two decades of radial velocity (RV) observations using Keck/HIRES and McDonald/Tull, and more recent observations using the Automated Planet Finder, we found that the nearby star HR 5183 (HD 120066) hosts a 3\mjup{} minimum mass planet with an orbital period of \period{} years. The orbit is highly eccentric (e$\simeq$0.84), shuttling the planet from within the orbit of Jupiter to beyond the orbit of Neptune. Our careful survey design enabled high cadence observations before, during, and after the planet's periastron passage, yielding precise orbital parameter constraints. We searched for stellar or planetary companions that could have excited the planet's eccentricity, but found no candidates, potentially implying that the perturber was ejected from the system. We did identify a bound stellar companion more than 15,000 au from the primary, but reasoned that it is currently too widely separated to have an appreciable effect on HR 5183 b. Because HR 5183 b's wide orbit takes it more than 30~au (1'') from its star, we also explored the potential of complimentary studies with direct imaging or stellar astrometry. We found that a Gaia detection is very likely, and that imaging at 10 $\mu$m is a promising avenue. This discovery highlights the value of long-baseline RV surveys for discovering and characterizing long-period, eccentric Jovian planets. This population may offer important insights into the dynamical evolution of planetary systems containing multiple massive planets.
\end{abstract}

\section{Introduction}

Radial velocity (RV) and transit surveys have characterized very few planets beyond 5 au \citep{Howard:2010aa, Mayor:2011aa, Petigura:2013aa, Fressin:2013aa}, leaving the population characteristics of long-period planets largely unknown. Direct imaging is sensitive to such planets, but the current generation of instruments is limited to planets several times the mass of Jupiter orbiting young, massive stars \citep{Bowler:2016aa, Nielsen:2019aa}. Microlensing is also sensitive to planets at large separations from their stars, and microlensing results already allow for occurrence calculations of planet mass as a function of separation \citep{Suzuki:2016aa}. However, mircolensing results will not enable detailed orbital or system architecture characterization. On the other hand, RV surveys are limited by their baselines. Several authors have used RV trends or other incomplete orbital arcs to constrain the properties of long-period planets and substellar objects \citep{Wright:2007aa, Wright:2009aa, Knutson:2014aa, Bouchy:2016aa, Rickman:2019aa, Bryan:2016aa}, but it is challenging to pin down the physical parameters of planets with orbital periods much longer than the survey baseline. Some authors assume circular orbits in order to cut down the wide parameter space of possible orbits \citep[e.g.][]{Knutson:2014aa}, but even so posteriors over semimajor axis and minimum mass span wide ranges. 

Long-baseline RV surveys dating back to the mid-1980s \citep{Campbell:1983aa, Marcy:1983aa, Mayor:1985aa, Campbell:1988aa, Marcy:1989aa, Zechmeister:2013aa, Fischer:2014aa, Wittenmyer:2014aa, Marmier:2013aa, Moutou:2015aa, Endl:2016aa} are beginning to fill this characterization gap as their time baselines increase. The long-period ($>1$ yr) planets discovered by these surveys share characteristics with the directly imaged planets and the shorter-period RV-discovered planets. As these surveys mature, they will allow us to characterize the transition from older, less massive, shorter-period RV-detected planets to younger, more massive, longer-period imaged planets. These new discoveries will also enable us to calculate the fundamental properties of planets in wider mass and age ranges than those currently accessible to direct imaging alone, examine the rarity of the Earth-Jupiter-Saturn architecture, and test giant planet formation theories \citep{Cumming:2008aa, Wittenmyer:2006aa, Wittenmyer:2011aa, Wittenmyer:2016aa}. 

Here, we present the discovery of HR 5183 b, a highly eccentric planet with a semimajor axis of \sma{} au orbiting a $V=6.3$ G0 star. HR 5183 has been monitored for more than 20 years as part of the California Planet Search at Keck/HIRES and the long-duration RV planet survey at McDonald Observatory. After over 10 years of relatively constant RV measurements, HR 5183 began rapidly accelerating. In 2018, the RV measurements flattened out and turned over, an event associated with the planet's periastron passage. As we discuss later in the paper, this periastron passage event was information-rich, and allowed precise constraints on the planet's orbital parameters even without RV coverage over the entire orbital period. With an orbital period of \period{} years, HR 5183 b is the longest-period planet with a well-constrained orbital period and minimum mass detected with the RV technique.

This paper is organized as follows: in Section \ref{sec:obs}, we present our RV measurements of HR 5183. In Section \ref{sec:star}, we provide precise estimates of the stellar parameters of HR 5183, and in Section \ref{sec:planet}, we characterize the planet HR 5183 b. In Section \ref{sec:system}, we describe an extremely widely-separated ($>$ 15,000 au) stellar companion to HD 5183, and present the results of searches for additional stellar and planetary companions. In Section \ref{sec:astrometry}, we discuss prospects for multi-method detection of HR 5183 b. In Section \ref{sec:discuss}, we relate HR 5183 b to other exoplanet systems, comment on formation scenarios, and conclude.

\section{High-Resolution Spectra}
\label{sec:obs}

We began Doppler monitoring of HR 5183 in 1997 at Keck/HIRES and in 1999 at McDonald/Tull. We have also monitored HR 5183 on the Automated Planet Finder (APF) with high cadence since its commissioning in 2013. The RVs from all three spectrographs are shown in Figure \ref{fig:multipanel} and tabulated in Table \ref{tb:rvs}.

\begin{deluxetable*}{cccccc}
\tablecaption{Radial Velocities and S-index values \label{tb:rvs}}
\tabletypesize{\footnotesize}
\tablehead{
  \colhead{Time} & 
  \colhead{RV} & 
  \colhead{RV Unc.} & 
  \colhead{Inst.} &
  \colhead{$S_\mathrm{HK}$\tablenotemark{b}} &
  \colhead{$S_\mathrm{HK}$ Unc.} \\
  \colhead{(BJD - 2440000)} & 
  \colhead{(m s$^{-1}$)} & 
  \colhead{(m s$^{-1}$)} & 
  \colhead{} &
  \colhead{} &
  \colhead{}
}
\startdata
$10463.1705$ & $-63.5$ & $1.09$ & HIRES \tablenotemark{a} & 0.14 & 0.01\\ 
$10547.042$ & $-67.93$ & $1.16$ & HIRES \tablenotemark{a} & 0.14 & 0.01\\ 
$10838.155$ & $-59.22$ & $1.08$ & HIRES \tablenotemark{a} & 0.14 & 0.01\\ 
$10954.9271$ & $-63.3$ & $1.61$ & HIRES \tablenotemark{a} & 0.14 & 0.01\\ 
$11200.1185$ & $-57.71$ & $1.24$ & HIRES \tablenotemark{a} & 0.14 & 0.01\\ 
$11213.9789$ & $-51.69$ & $9.03$ & TULL & 0.15 & 0.02\\ 
$11241.8948$ & $-35.6$ & $4.08$ & TULL & 0.15 & 0.02\\ 
$11274.8541$ & $-49.01$ & $8.4$ & TULL & 0.16 & 0.02\\ 
$11310.9479$ & $-63.84$ & $1.34$ & HIRES \tablenotemark{a} & 0.14 & 0.01\\ 
$11329.7947$ & $-36.45$ & $4.72$ & TULL & 0.15 & 0.02\\ 
\enddata
\tablenotetext{a}{Pre-upgrade HIRES measurement.}
\tablenotetext{b}{Note that the $S_\mathrm{HK}$ values for each instrument do not have the same zero-point. Pre- and post-upgrade HIRES S-values should be treated independently.}
\tablecomments{Table \ref{tb:rvs} is published in its entirety in the machine-readable format. A portion is shown here for guidance regarding its form and content.}
\end{deluxetable*}

\subsection{HIRES Spectra}
\label{sec:hires}
We obtained \nobshires{} high-resolution ($R=60,000$) spectra of HR 5183 with the HIRES spectrograph \citep{Vogt:1994aa, Cumming:2008aa, Howard:2010aa} between 1997 and 2019. HIRES underwent major upgrades in 2004, so for modeling purposes we treat pre- and post-upgrade HIRES measurements independently (see Section \ref{sec:planet}). Wavelength calibration for each RV measurement was performed with a warm iodine-gas cell placed in the light path in front of the slit, producing a convolved spectrum of the star, iodine gas, and point spread function. Each spectrum was forward-modeled with a deconvolved stellar spectrum template (DSST), an atlas iodine spectrum, and a line spread function \citep{Butler:1996aa}. This technique is stable at the 2-3 $m s^{-1}$ level on timescales of more than a decade \citep{Howard:2016aa}.

% The HIRES spectrograph is mounted at one Nasmyth focus on the Keck I 10m telescope. It has optical wavelength coverage from 365-800 nm across three charge coupled devices (CCDs). The spectral resolution at 550 nm, in the middle of the iodine region, is $R=60,000$. 

% At $V=6.3$, HR 5183 is bright enough that we often observed it at bright times, including before the dusk 12 degree twilight and after the dawn 12 degree twilight. 

To monitor chromospheric and stellar spot activity, we extracted spectral information at and near the Ca II H and K lines to calculate a Mt. Wilson style S-index value (following \citealt{Wright:2004aa} and \citealt{Isaacson:2010aa}) for measurements taken after the 2004 instrument upgrade. S-index values for HIRES measurements taken before 2004 were pulled directly from \citet{Wright:2004aa}. These values do not correlate significantly with time, the RV measurements, or the RV residuals from the maximum a posteriori (MAP) orbit (see Section \ref{sec:planet}). In particular, the S-index values show no trends or correlations with RV measurements on the timescale of the proposed planet period. 

\subsection{Tull Spectra}
Between 1999 and 2019, we collected  \nobsmcd{} high-resolution ($R=60,000$) spectra with the Tull Coud\'e Spectrograph \citep{Tull:1995aa} on the 2.7\,m Harlan J. Smith telescope as part of the McDonald Observatory planet search \citep{Cochran:1997aa, Hatzes:2000aa}. For all observations, we inserted an iodine absorption cell into the light path to obtain a precise wavelength calibration. Combined with a template stellar spectrum, this allowed us to reconstruct the shape of the instrumental PSF at the time of each observation. We used the RV modeling code \texttt{Austral} \citep{Endl:2000aa} to compute precise differential RVs. 

We typically reach a long-term RV precision of 4 to 6 \ms{} for inactive FGK-type stars with the Tull spectrograph. A major advantage of the Tull RV survey is that the instrumental setup has not been modified over the duration of the program. For nearly 20 years, we have been using the same CCD detector, the same iodine cell, and the same positions of the Echelle grating and cross-disperser prism. This assures that there are no RV zero-point offsets introduced into the RV time series.

We determined the S-index values from the Ca II H\&K lines in the blue orders of the Tull spectra using the method outlined in \citet{Paulson:2002aa}. These S-index values also show no trend or correlation with RV measurements over the duration of the observations.

\subsection{APF Spectra}
Finally, we obtained \nobsapf{} spectra of HR 5183 with the Automated Planet Finder (APF; \citealt{Radovan:2014aa, Vogt:2014aa}) between 2013 and 2019. The APF is an automated 2.4 meter telescope at Lick Observatory on Mt. Hamilton, CA. It is equipped with the Levy Spectrograph, a dedicated high-resolution echelle spectrometer that sits at a Nasmyth focus. The Levy Spectrograph achieves $ R > 120,000$ and covers a wavelength range of 374.3-980.0 nm. Spectra of HR 5183 were observed through a warm iodine-gas cell for wavelength calibration. The RVs were calculated with the pipeline described in \citet{Fulton:2015aa}, which descends from the \citet{Butler:1996aa} pipeline, and is essentially identical to the HIRES reduction pipeline discussed in Section \ref{sec:hires}. As with the HIRES data, we calculate S-index values following \citet{Isaacson:2010aa}. These S-index values similarly appear independent of the RV measurements over the duration of the observations.

\begin{figure*}[h]
    \centering
    \includegraphics[scale=.9]{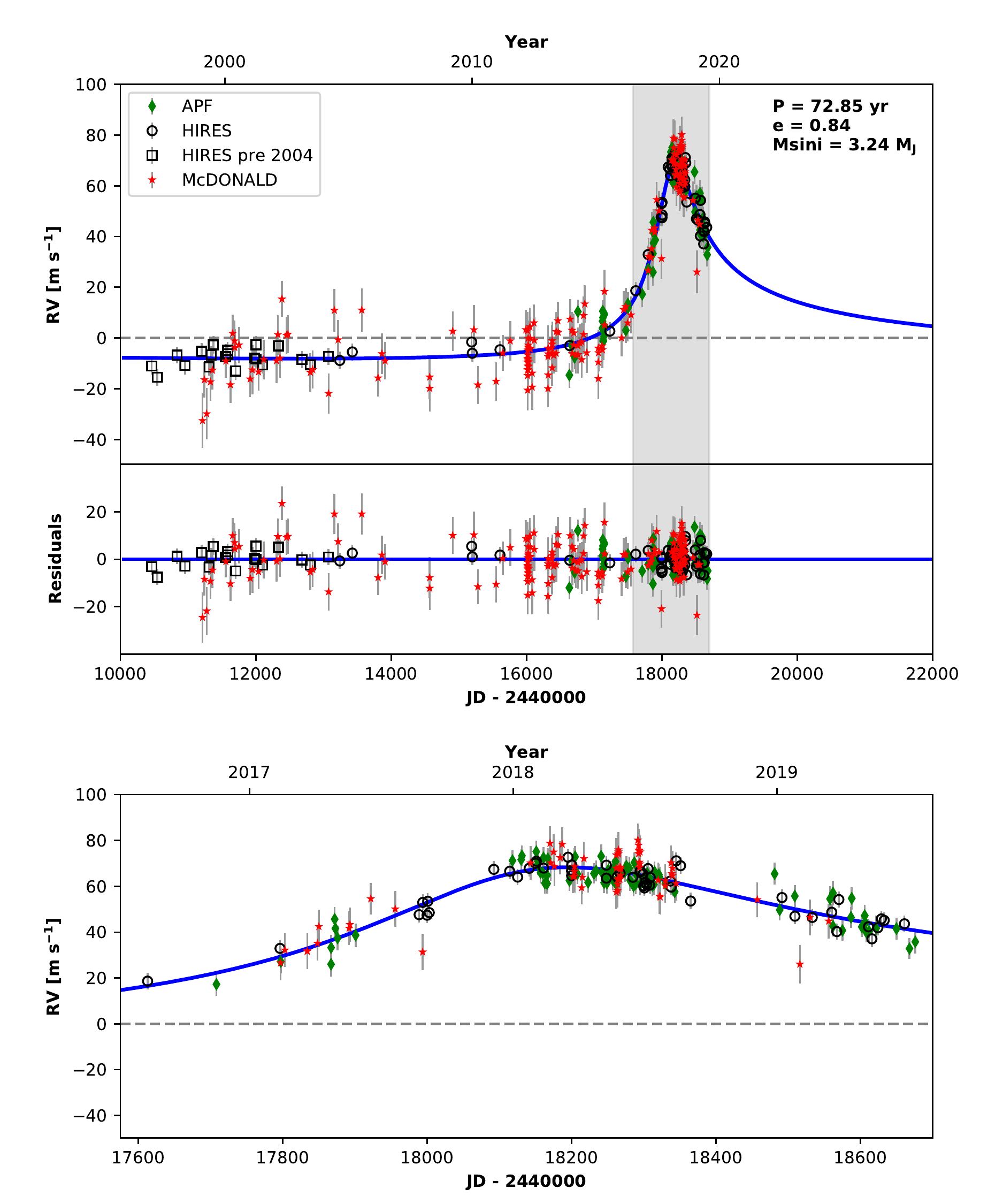}
    \caption{Top: RVs of HR 5183 from the Keck-HIRES, McDonald-Tull, and APF-Levy spectrographs as a function of time. Error bars show observational errors and instrument-specific jitter values added in quadrature. The best-fit Keplerian orbit is shown (blue solid line). Residuals are inset below. Bottom: close-up of the grey region in top plot. The RV curve peaks in January 2018 during periastron passage, and declines monotonically afterward in all three data sets.}
    \label{fig:multipanel}
\end{figure*}

\section{Stellar Properties}
\label{sec:star}

HR 5183 is a nearby slightly evolved G0 star. We derived precise stellar parameters for HR 5183 using the method described in \citet{Fulton:2018aa}. Briefly, this method uses Gaia DR2 parallaxes \citep{Gaia-Collaboration:2018aa}, spectroscopic effective temperatures computed from our Keck template spectrum with the \texttt{SpecMatch} code \citep{Petigura:2015aa}, and 2MASS photometry \citep{Skrutskie:2006aa} to compute precise stellar radii. $\log{g}$, $[\mathrm{Fe/H}]$, and $v \sin{i}$ are also calculated from the Keck spectrum using \texttt{SpecMatch}. Stellar  mass, age, and distance are derived using the \texttt{isoclassify}\footnote{GitHub.com/danxhuber/isoclassify} package \citep{Huber:2017aa}. The stellar properties derived from this analysis are presented in Table \ref{tb:st_props}, along with other useful stellar parameters. 

\citet{Allen:2014aa} found evidence that HR 5183 is in the halo of the Milky Way using reduced proper motion diagrams following \citet{Salim:2003aa}. However, HR 5183 is younger and more metal-rich than typical galactic halo objects \citep{Carollo:2016aa}, which led us to scrutinize this claim. To investigate HD 5183's galactic population membership, we performed a kinematic analysis of its galactic orbit, following \citet{Johnson:2018aa}. We used the \texttt{galpy}\footnote{GitHub.com/jobovy/galpy} package \citep{Bovy:2015aa} to compute 50 random realizations of galactic positions and U,V,W space velocities for HR 5183 consistent with its Gaia DR2 parameters \citep{Gaia-Collaboration:2018aa}. For each realization, we then calculated the galactic orbit of HR 5138 in \texttt{galpy}'s ``MWPotential2014'' galactic potential. The resulting orbits never achieve a height above the galactic midplane of more than 200 pc. This result supports the claim that HR 5183 is a thin-disk member, and not a halo object. 

\begin{deluxetable*}{ccc}
\tablecaption{Stellar Properties \label{tb:st_props}}
\tabletypesize{\footnotesize}
\tablehead{
  \colhead{Parameter} & 
  \colhead{Value} & 
  \colhead{Unit}
}
\startdata
R.A. & 13 46 57 & hh:mm:ss \\
Decl.& +06 20 59 & dd:mm:ss \\ 
HD Name & HD 120066 & ---\xspace \\
2MASS ID & J13465711+0621013 & ---\xspace \\
Gaia Source ID & 3721126409323324416 & ---\xspace \\
Parallax & $31.757\pm0.039$ & mas \\
$K$ & $4.85\pm0.02$ & mag \\
$V$ & $6.30$ & mag \\
$T_{\mathrm{eff}}$ & $5794\pm100$ & K \\
$\log{g}$ & $4.02\pm0.1$ & dex \\
$[\mathrm{Fe/H}]$ & $0.10\pm0.06$ & dex \\
$v \sin i$ & $3\pm1$ & $km s^{-1}$ \\
$R_*$ & $1.53^{+0.06}_{-0.05}$ & $R_\odot$ \\
$M_*$ & $1.07\pm0.04$ & $M_\odot$ \\
Age & $7.7^{+1.4}_{-1.2}$ & Gyr \\
Distance & $31.49\pm0.04$ & pc \\
\enddata
\tablecomments{$T_{\mathrm{eff}}$, $\log{g}$, $[\mathrm{Fe/H}]$, and $v \sin i$ were calculated from the stellar spectrum using the \texttt{SpecMatch} code. $R_*$ was calculated as described in Section \ref{sec:star}. $M_*$, age, and distance were calculated using the \texttt{isoclassify} code.}
\end{deluxetable*}

\section{Planet Properties}
\label{sec:planet}

The curvature we saw in the RVs (see Figure \ref{fig:multipanel}) alerted us to the existence of HR 5138 b, and motivated us to characterize its orbital properties. We modeled the RV timeseries using the open-source toolkit \texttt{radvel}\footnote{https://radvel.readthedocs.io/en/latest} \citep{Fulton:2018aa}. The code and data used to perform the analysis in this paper are available on GitHub\footnote{https://github.com/California-Planet-Search/planet-pi}. We chose to perform this fit using the following parametrization of the Keplerian RV function: {$\log{P}$, T$_C$, $\sqrt{e}\cos{\omega}$, $\sqrt{e}\sin{\omega}$, and $\log{K}$}. We imposed uninformative uniform priors on each of these parameters except $\log{P}$, for which we defined an ``informative baseline prior.'' Because we detected a long-period planet by observing a single, short-duration event (the planet's periastron passage), we made an analogy to detection by transit and defined the following prior on period, often used in the exoplanet transit community \citep[e.g.,][]{Kipping:2018aa, Vanderburg:2016aa}:

\begin{equation}
    p(P, t_d, B) = 
    \begin{cases}
        1 & \mathrm{if} \quad P-t_d < B \\
        (B+t_d)/P & \mathrm{else}
    \end{cases}
\end{equation} where $t_d$ is the duration of the event (in this case, the periastron passage), $P$ is the orbital period, and $B$ is the observing baseline. See Section \ref{sec:model-sel} for a justification of this choice of prior and detailed comparison to other possible models and prior parameterizations. Unlike in the case of a transit detection, the ``duration'' of HR 5183's periastron passage event is not easily defined. We performed fits with $t_d=0$ and $t_d=3.5$ yr, ultimately finding that the results were indistinguishable and sidestepping this issue. We adopted $t_d=0$ for convenience. 

We also included jitter ($\sigma$) and RV offset ($\gamma$) terms for each of our four RV datasets (we treated HIRES pre-2004 and post-2004 measurements as separate data sets in our fit; see Section \ref{sec:hires}). We assumed uninformative uniform priors on each of these instrumental terms as well. The logarithm of the complete likelihood for this model is:

\begin{equation}
    \ln\mathcal{L} = -\frac{1}{2} \sum^n_{i=0} \left[ \frac{(v_i - M_i)^2}{(\sigma_i + \sigma_{\mathrm{jit},i})^2} + 2\mathrm{ln}\sqrt{2\pi (\sigma_i^2 + \sigma_{\mathrm{jit},i}^2)^{\frac{1}{2}}}\right]
\end{equation}
where n is the total number of RV measurements, $v_i$ is the $i$th RV measurement, $\sigma_i$ is its uncertainty, $M_i$ is the Keplerian model prediction for observation $i$, and $\sigma_{\mathrm{jit},i}$ is the jitter parameter for the instrument that took observation $i$ \citep{Fulton:2016aa}. 

\begin{figure}[h!]
    \centering
    \includegraphics[width=0.48\textwidth]{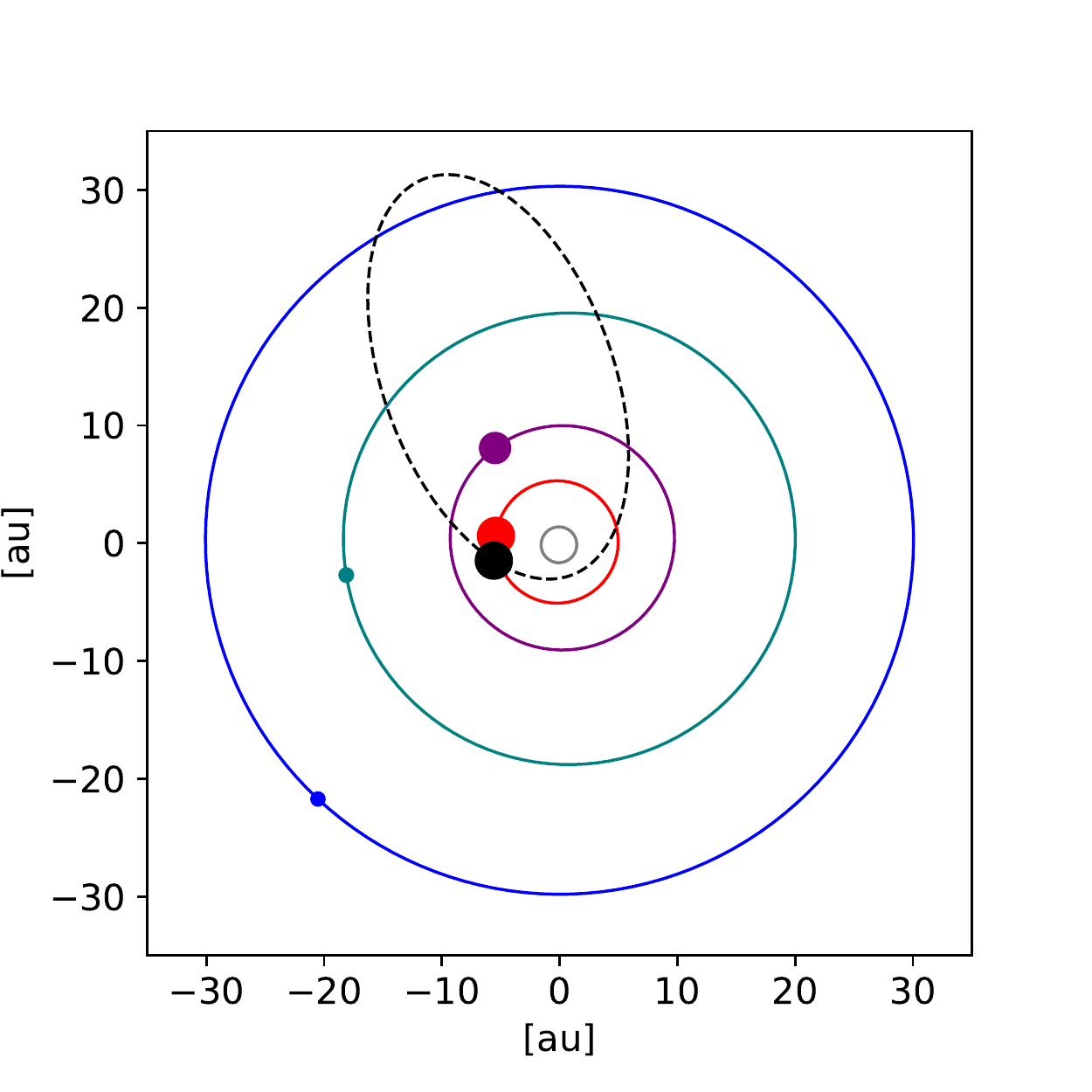}
    \caption{The MAP orbit of HR 5183 b compared to the orbits of the planets in our solar system (blue: Neptune, teal: Uranus, purple: Saturn, red: Jupiter, grey: Mars, black dashed: HR 5183 b). The sizes of the colored circles show the relative sizes of the solar system planets (not to scale with respect to their orbits). We assume a radius of 1 R$_J$ for HR 5183 b. The orbital locations of the planets are computed on 7-31-2019, approximately 1.5 years after the periastron passage of HR 5183 b. HR 5183 b's $\Omega$ and $i$ are set to arbitrary values. At periastron, HR 5183 b is closer to its star than our asteroid belt is to the Sun, and at apastron, it is more distant than Neptune.}
    \label{fig:solar_system}
\end{figure}

\begin{figure*}[h!]
    \centering
    \includegraphics[scale=0.45]{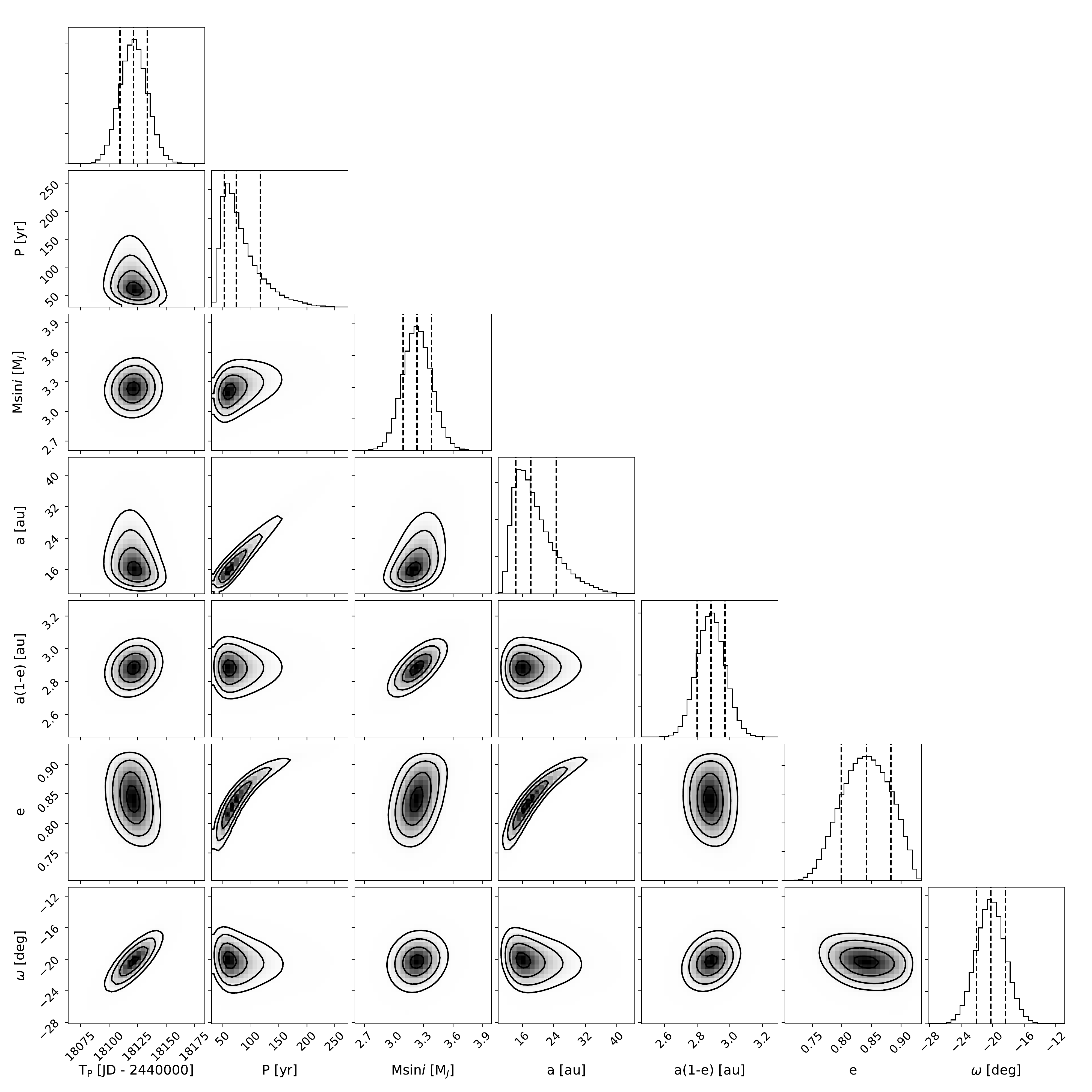}
    \caption{Corner plot displaying posterior distributions and covariances of HR 5183 b orbital parameters of interest: time of periastron passage (T$_P$), orbital period ($P$), minimum mass ($M\sin{i}$), semimajor axis ($a$), minimum orbital separation ($a(1-e)$), eccentricity ($e$), and argument of periastron ($\omega)$. Importantly, these probability distributions were derived from the fitted posteriors; among the parameters shown here, only T$_P$ is actually a parameter in the fit. Time of periastron passage, minimum mass and minimum orbital separation are very well constrained, while orbital period remains uncertain and correlated with eccentricity.}         
    \label{fig:derived}
\end{figure*}

We computed the MAP fit with \texttt{radvel}, obtaining an orbital period of \mapperiod{} years, a minimum mass of \mapmsini{} $M_J$, and an eccentricity of \mapecc{}. This orbital solution is shown in Figure \ref{fig:multipanel}, and a bird's-eye view comparing this orbit to the orbits of the solar system planets is shown in Figure \ref{fig:solar_system}. We next performed an Affine-invariant Markov Chain Monte Carlo (MCMC) exploration of the parameter space with the ensemble sampler \texttt{emcee}\footnote{GitHub.com/dfm/emcee} \citep{Foreman-Mackey:2013aa}. Our MCMC analysis used \nensembles{} ensembles of \nwalkers{} walkers and ran for \nsteps{} steps per walker, achieving a maximum Gelman-Rubin (GR; \citealt{Gelman:2003aa}) statistic of \maxGR{}. A corner plot showing posterior distributions and covariances between $T_P$, $P$, $M\sin{i}$, $a$, $a(1-e)$, $e$, and $\omega$ is shown in Figure \ref{fig:derived}. These values are also recorded in Table \ref{tb:pl_props}. 

\subsection{Model Choice}
\label{sec:model-sel}

We performed three additional orbit fits to evaluate our choice of model. First, we performed two fits without informative baseline priors on $\log{P}$: one fitting in $P$ and $K$ (as opposed to $\log{P}$ and $\log{K}$), and another in $\log{P}$ (fitting in $\log{P}$ versus $P$ imposes an implicit Jeffrey's prior on $P$). Period posteriors obtained from these two models and the informative prior are compiled in Table \ref{tb:prior_comp}. All three orbital period posteriors are consistent within 1-$\sigma$, but the informative baseline prior pushes the median orbital period to shorter values. Neither of these priors significantly change the posteriors on the other orbital parameters; for example, fitting in $\log{P}$ gives $M\sin{i}$ = \msinilogP{} \mjup{} and e = \ecclogP{}. The slight dependence of the solution on our prior choice ultimately points to the need for more data, but in the meantime we adopt the informative prior. 

Next, we performed a fit including a $\dot{\gamma}$ parameter to account for potential additional wider-separation companions influencing the RV signal of the star. Adding this free parameter has the effect of pushing the eccentricity posterior to higher values (e = \gammadotecc{}) and the period posterior to lower values (P = \gammadotper{} yr), but the posterior distribution of $\dot{\gamma}$ is consistent with 0 ($\dot\gamma$ = \gammadot{} m s$^{-1}$ yr$^{-1}$). The adopted model has lower BIC ($\Delta$BIC = \deltabic{}) and AIC values ($\Delta$AIC = \deltaaic{}), indicating that the added free parameter does not substantially improve the fit. We can't unequivocally rule out a trend, but since including one is not statistically warranted and does not affect the conclusions of the paper, we adopt the fit with no trend. 

The lack of an unambiguous trend in the RVs is consistent with our failure to detect companion objects in the HST and NaCo images (see Section \ref{sec:star} and Appendix \ref{sec:images}). The possible bound companion at 15,000 au (Section \ref{sec:stellar_comp}) would not produce a measurable $\dot{\gamma}$.

\begin{deluxetable*}{cccc}
\tablecaption{Fit Parameters \& Derived Planet Properties \label{tb:pl_props}}
\tabletypesize{\footnotesize}
\tablehead{
  \colhead{Parameter} & 
  \colhead{Median Value \& 68\% CI} & 
  \colhead{MAP Value} & 
  \colhead{Unit}
}
\startdata
$\ln P$ & \logper & \maplogper & ln(days) \\
$T_{\mathrm{c}}$ & \tconj & \maptc & JD - 2440000 \\
$\sqrt{e}\cos{\omega}$ & \secosw & \mapsecosw & \\
$\sqrt{e}\sin{\omega}$ & \sesinw & \mapsesinw &\\
$\ln K$ & \logK & \maplogK & ln(\ms{}) \\
$\sigma$ (HIRES pre-upgrade) & \jitK & \mapjitK & \ms{} \\
$\sigma$ (HIRES post-upgrade) & \jitJ & \mapjitJ &\ms{} \\
$\sigma$ (Tull) & \jitM & \mapjitM &\ms{} \\
$\sigma$ (APF) & \jitA & \mapjitA &\ms{} \\
$\gamma$ (HIRES pre-upgrade) & \gammaK & \mapgammaK &\ms{} \\
$\gamma$ (HIRES post-upgrade) & \gammaJ & \mapgammaJ & \ms{} \\
$\gamma$ (Tull) & \gammaM & \mapgammaM & \ms{} \\
$\gamma$ (APF) & \gammaA &  \mapgammaA & \ms{} \\
\hline
$P$ & \period & \mapperiod & yr \\
$K$ & \semiamp & \mapK  & ms$^{-1}$ \\
$e$ & \ecc & \mapecc  & \\
$\omega$ & \aop &  \mapaop  & rad  \\
$T_P$ & \tp &  \mapTp  & JD - 2440000 \\
$M \sin{i}$ & \msini &  \mapmsini  &$M_J$ \\
$a$ & \sma &  \mapa  & au \\
$a(1-e)$ & \closestapproach &  \mapclosestapproach  &au \\
$T_{\mathrm{eq}}$ (peri) & \teqperi & \mapteqperi & K \\
$T_{\mathrm{eq}}$ (apo) & \teqapo &  \mapteqapo  & K \\
\enddata
\tablecomments{$T_{\mathrm{eq}}$ values were calculated assuming a visible albedo of 0.5.}
\tablecomments{$\omega$ refers to the orbit of the star HR 5183 induced by the planet HR 5183 b.}
\end{deluxetable*}

\begin{deluxetable*}{cc}
\tablecaption{Period Prior Comparison \label{tb:prior_comp}}
\tabletypesize{\footnotesize}
\tablehead{
  \colhead{Prior} & 
  \colhead{Median Period \& 68\% CI} 
}
\startdata
uniform in $P$ \& $K$ & $125^{+113}_{-54}$ yr \\uniform in $\log P$ \& $\log K$ & $103^{+103}_{-41}$ yr \\inf. baseline prior (adopted) & $74^{+43}_{-22}$ yr
\enddata
\end{deluxetable*}

\subsection{Orbit Information Density}

Our measurements of this planet's properties may seem surprisingly precise (see Table \ref{tb:pl_props}) given observations spanning only about one third of the orbit. These constraints are possible because we tracked the system through periastron passage, when the information density of the Keplerian signal is highest.  

High-eccentricity orbits have unique shapes that sensitively depend on $e$ and $\omega$ (see Fig.\ 2 of \citealt{Howard:2016aa} for a helpful visualization). The shape of the HR 5183 RV curve is fit only by a narrow range of these parameters, as Figure \ref{fig:derived} shows. 

The relatively flat RV curve from $\sim$1998--2015 followed by a sharp uptick and subsequent turnover are consistent only with $e\simeq0.8$ and $\omega\simeq-0.4$. All other Keplerian curves have \textit{shapes} that are inconsistent with our measurements. More complicated models involving additional planets or a $\dot\gamma$ term are also excluded by the peculiar RV pattern.

We offer two arguments to build intuition. First, imagine decomposing the RV fitting into a process that matches three orbital properties of the Keplerian curve: 1) the shape (from $e$ and $\omega$); 2) the vertical scale ($K$); and 3) the horizontal scale ($P$). Once the shape has been determined by matching the appropriate Keplerian curve, the horizontal and vertical scales can be measured using RVs spanning less than a full orbit, provided the information-rich close approach is covered.  
Second, consider the how the planet's speed varies over its orbit.  
We can define the ``fastest half orbit'' as the portion of an orbit near closest approach, when the true anomaly ($f$) is between $-\pi/2$ and $\pi/2$. The time for the planet to pass through the fastest half orbit, $t_\mathrm{fho}$, can be computed using the relationship between $f$ and time ($t$), 

\begin{equation}
    df = \frac{2\pi}{P} \sqrt{1-e^2}\left(\frac{a}{r}\right)^2 dt,
\end{equation}
where $r$ is the distance between the orbiting planet and the star \citep[][Eq.\ 2.44]{Seager:2010aa}. Substituting an expression for $r(f)$ \citep[][Eq.\ 2.20]{Seager:2010aa}, 

\begin{equation}
    \label{eq:peri}
    t_\mathrm{fho} = \frac{P(1-e^2)^{3/2}}{2\pi} \int_{-\pi/2}^{\pi/2}\frac{df}{1+e\cos{f}}.
\end{equation} 

For a circular orbit, $t_\mathrm{fho}$ integrates to $P/2$, as expected. Eccentric orbits have much shorter timescales of close approach though. Numerically integrating Eq.\ \ref{eq:peri} with $e = 0.84$, we find $t_\mathrm{fho} \approx P/18.6$. That is, the planet completes the fastest half of its orbit nearly an order of magnitude more quickly than in the circular case. While our fitting procedure did not actually measure $t_\mathrm{fho}$ and scale it by a factor of 18.6 to determine $P$, this exercise illustrates how highly eccentric orbits contain information related to orbital period on short timescales, and thus allow us to measure $P$ with higher precision than one might expect.

This is not to say that we have ruled out hundred-year or more periods and higher ($\simeq$0.91) eccentricities. Such orbits appear in our posterior, but because there is less posterior volume in this region of parameter space, they are less probable overall. 

\section{Additional Bound Companions}
\label{sec:system}

\subsection{Search for Additional Planets in the System}

We searched for other significant periodic signals in the RV data using the $\chi^2$ difference technique described in \citet{Howard:2016aa}. In brief, we started by calculating the $\chi^2$ of a flat line fit to the RVs, and injecting additional Keplerian orbits into the model. We calculated the change in $\chi^2$ ($\Delta\chi^2$) when including each additional Keplerian orbit over a grid of periods and eccentricities. We constructed a periodogram of the $\Delta\chi^2$ values as a function of trial period and fit the distribution of periodogram peak heights to infer an empirical false alarm probability (eFAP) for each detected peak. We detected no signals with an empirical false-alarm probability (eFAP) greater than 1\%, indicating no additional planetary companions down to our sensitivity limits.

\begin{figure}[h!]
    \centering
    \includegraphics[width=0.48\textwidth]{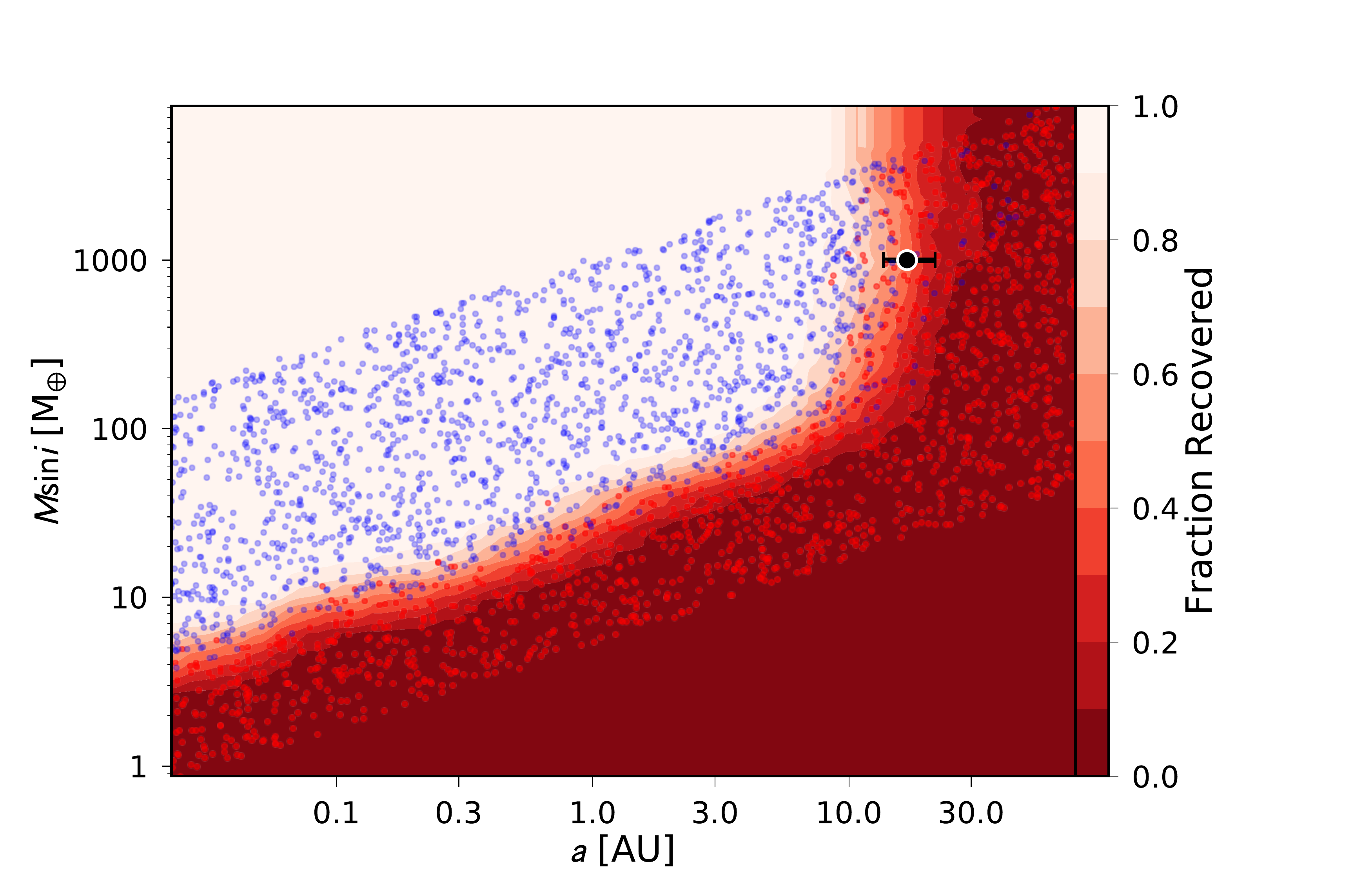}
    \caption{Sensitivity of the HR 5183 RV time series to injected planetary signals as a function of semimajor axis ($a$) and minimum mass ($M\sin{i}$). Each point corresponds to an injected radial velocity signal. Blue dots were detected, while red dots were not. The solid-color background shows the fraction of signals that were recovered, corresponding to the probability of detection. The parameters of HR 5183 b are shown as a black point with error bars (the uncertainty on $M\sin{i}$ is too small see). Our data are sensitive to less massive, shorter-period planets out to the orbit of HR 5183 b. }
    \label{fig:rv_search}
\end{figure}

We characterized our sensitivity limits over a grid of semimajor axes and $M\sin{i}$ values by applying the search algorithm described above to each injected planetary signal. The results of this analysis are shown in Figure \ref{fig:rv_search}. As expected, we are most sensitive to Jupiter-mass and heavier planets with $a<30$ au. Our data are not sensitive to Earth-mass planets. HR 5183 b itself is at our detection limits because of its large semimajor axis, but its high eccentricity makes it detectable.

We also searched for transit signals in ground-based photometric observations of HR 5183, finding no significant signals above our sensitivity limits. These data and analysis are described in Appendix \ref{sec:photom}. 

\subsection{Search for Stellar Companions to HR 5138}

\begin{figure}[htbp]
    \begin{center}
    \includegraphics[width=0.48\textwidth]{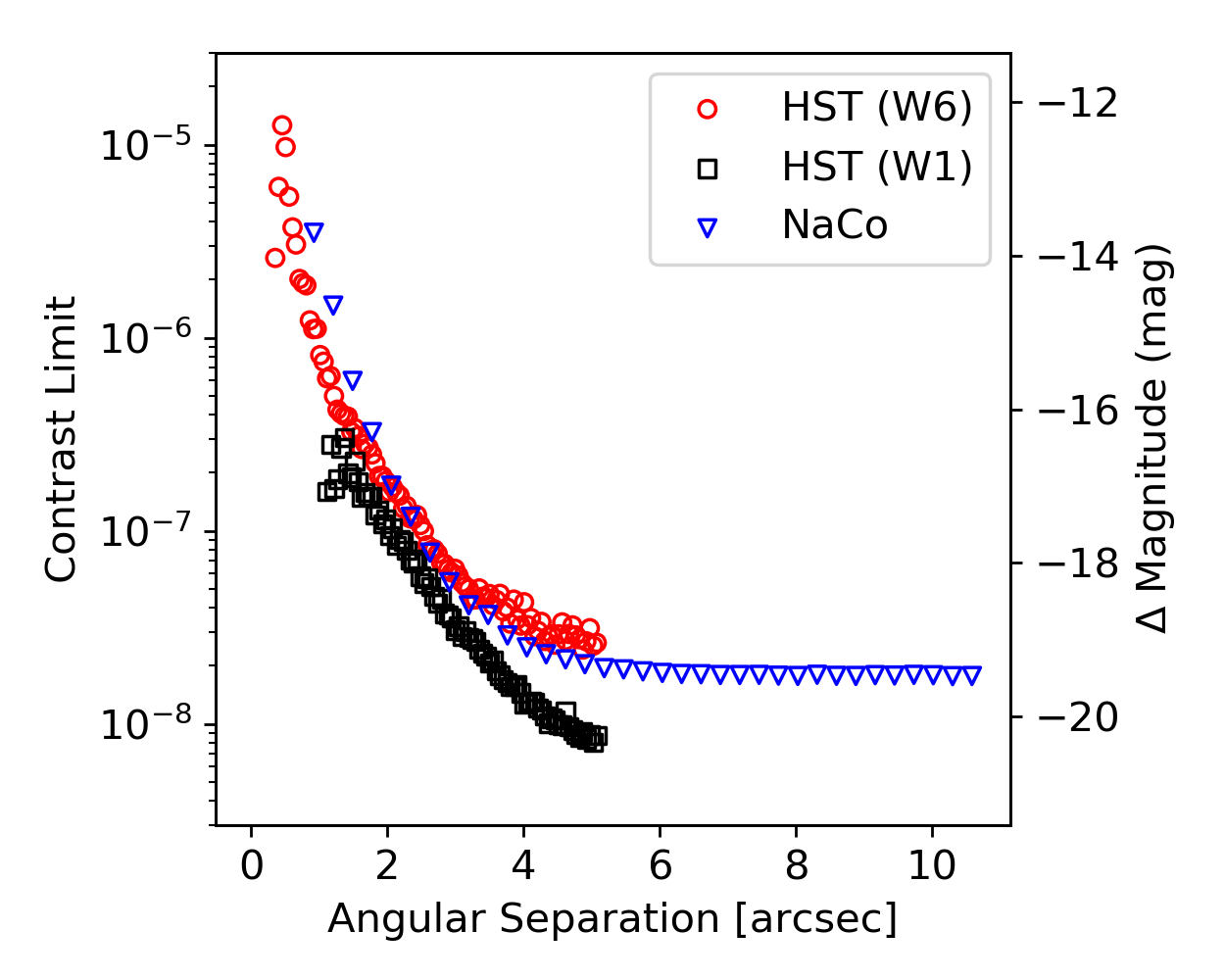}
    \caption{\label{fig:images} 1-$\sigma$ point-source detection limits for HR 5183 computed with NaCo on the VLT in the Ks-band, and with HST-STIS using WedgeA-0.6 (W6) and WedgeA-1.0 (W1). See Appendix \ref{sec:images} for details. We did not detect companions to HR 5183 in these images.}
    \end{center}
\end{figure}

We used a two-pronged approach to search for additional bound companions to HR 5183: analyzing archival coronagraphic images of the star and searching the Gaia DR2 database for stars with similar 3D locations and kinematic properties. HR 5138 b is likely much below the detection limit of current coronagraphic imagers (see Section \ref{sec:astrometry}), and we did not expect to detect it in these images. We found several archival images of HR 5183: one set of images taken with VLT's Nasmyth Adaptive Optics System (NAOS) Near-Infrared Imager and Spectrograph (CONICA), hereafter NaCo, and one set taken with the Hubble Space Telescope Imaging Spectrograph (HST/STIS). Details about the observations and data reduction are presented in Appendix \ref{sec:images}. We used these images to derive contrast curves illustrating our detection limits for HR 5183 (Figure \ref{fig:images}), and found no evidence for companions, with sensitivity down to $\Delta$mag = 20 at 4".

While our scrutiny of coronagraphic images revealed no companions, through our Gaia DR2 search and the analysis described below, we found that HIP 67291 is likely an eccentric, widely separated ($>$ 15,000 au) stellar companion to HR 5138. However, even if this star is gravitationally bound to HR 5183, it is too widely separated to affect the planet HR 5183 b. In addition, it would not be in the field of view of any of the images described in Appendix \ref{sec:images}.

\begin{figure*}[h]
    \centering
    \includegraphics[width=\textwidth]{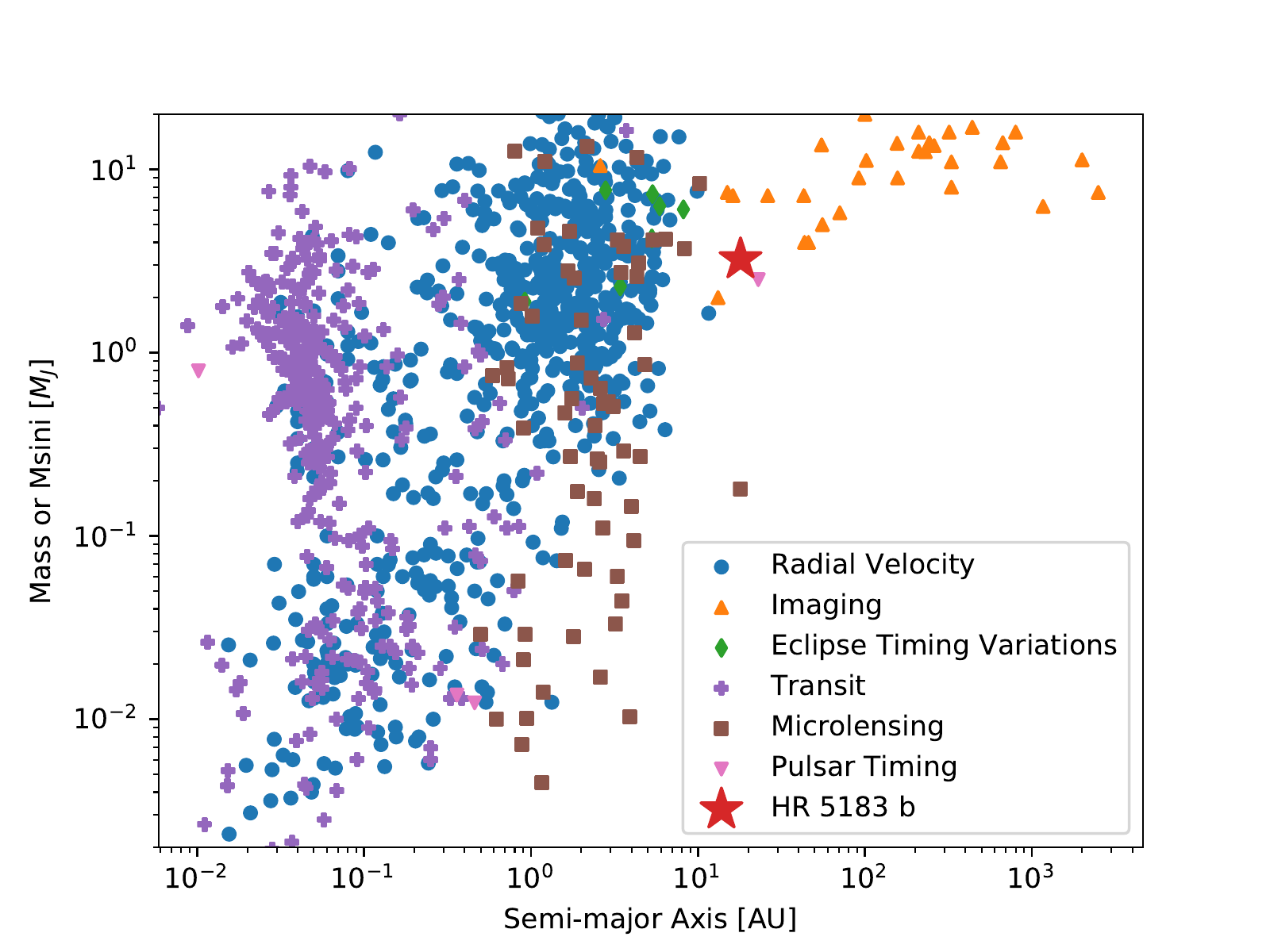}
    \caption{Mass or $M$sin$i$ versus semimajor axis for planets in the NASA Exoplanet Archive (accessed 2019-4-14; colored based on detection method) and HR 5183 b (1-$\sigma$ error bars; black square). Planets discovered with the RV or transit techniques are only included if their orbital periods are 1.5-$\sigma$ detections or better. HR 5183 b straddles the boundary between radial velocity and directly imaged exoplanets in this space, allowing us to probe the transition between the two populations.}
    \label{fig:long-period-planets}
\end{figure*}

\subsubsection{HIP 67291: A Wide Stellar Companion to HR 5183}
\label{sec:stellar_comp}

Several papers in the literature have presented evidence that HIP 67291, a K7V star \citep{Alonso-Floriano:2015aa} with a projected separation of more than 15,000 au, is bound to HR 5183 (\citealt{Allen:2000aa}, \citealt{Tokovinin:2014aa}, \citealt{Allen:2014aa}). Using kinematic parameters from Gaia DR2 and an isochrone-derived mass for HIP 67291, we investigated the probability that these two stars are gravitationally bound, and present orbital parameters for the system. This analysis is meant to be exploratory and not definitive; additional undetected companions orbiting HIP 67291 would affect these calculations, for example.

We performed an isochrone fit for HIP 67291 using the \texttt{isochrones}\footnote{GitHub.com/timothydmorton/isochrones} Python package \citep{Morton:2015aa} to interface with the MIST stellar evolution models \citep{Dotter:2016aa, Choi:2016aa, Paxton:2011aa, Paxton:2013aa, Paxton:2015aa}. We defined priors on [Fe/H] and $\log{g}$ using the values and precisions used in the template to compute the Gaia radial velocity of HIP 67291 (\texttt{rv\_template\_fe\_h} and \texttt{rv\_template\_logg} in the Gaia DR2 database, respectively). In addition, we placed Gaussian priors on parallax and $T_{\mathrm{eff}}$, informed by the Gaia DR2 values and uncertainties reported for HIP 67291. We also placed a Gaussian prior on the age of HIP 67291, informed by the age of HR 5183 derived in Section \ref{sec:star}, but found that this constraint did not affect the mass of HIP 67291. We obtained a mass of $0.67\pm0.05$ \msun{} from this analysis, which is consistent with the K7V spectral type derived in \citet{Alonso-Floriano:2015aa}.

\begin{figure}[h]
    \centering
    \includegraphics[scale=0.75]{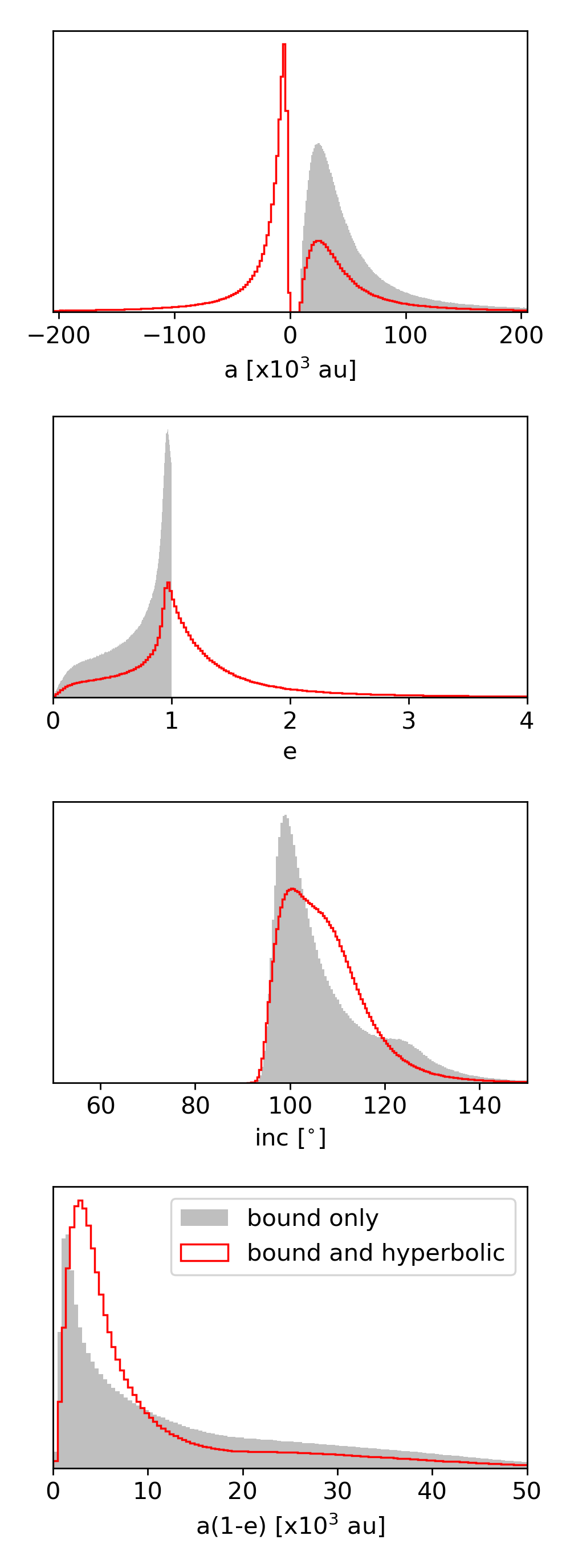}
    \caption{Bound (grey solid) and all (red line) solutions for the orbit of HIP 67291 and HR 5183. These are normalized, so even though the grey PDF is a subset of the red PDF, the grey exceeds the red in places. These solutions may not be accurate if there are undetected massive companions around HIP 67291. High eccentricities and edge-on orbits are preferred. Although small values of a(1-e) (neglecting unbound hyperbolic orbits) are possible, the most probable value occurs at 1,000 au, and the median at 10,000 au, too widely separated to affect the planet HR 5183 b. The most probable orbital period is almost 1 Myr.}
    \label{fig:companion}
\end{figure}

Given the mass of HIP 67291, the mass of HR 5183 derived in Section \ref{sec:star}, and the respective parallaxes, R.A./Dec.\ values, proper motions, and radial velocities of both stars from Gaia DR2 (compiled in Table \ref{tb:gaia}, the orbit of the two stars is in principle completely specified. In practice, the uncertainties on these parameters are significant enough to permit large uncertainties in the orbital parameters. To quantify these uncertainties, we drew samples from Gaussian distributions over both stellar masses and each of the six positional and velocity measurements for each star, then calculated the resulting orbital parameters. We found that 44\% of these generated orbits had $e<1$ (i.e. are bound). Histograms of the orbital parameters derived from this sampling method (the likelihood over possible bound and unbound/hyperbolic orbital parameters) are shown in Figure \ref{fig:companion}. Highly eccentric, edge-on orbits are preferred. 

While the likelihood that these two stars are bound is only 44\%, the two possible physical explanations for the 66\% of hyperbolic orbits (that the two stars are currently ``flying-by'' one another and that they were bound in the past and recently became unbound) likely have low prior probabilities. Therefore, the posterior probability that the two stars are bound is likely much higher than 44\%.

\begin{deluxetable*}{ccccccc}
\tablecaption{Gaia DR2 Parameters for HR 5183 and HIP 67291 \label{tb:gaia}}
\tabletypesize{\footnotesize}
\tablehead{
  \colhead{Parameter} &
  \colhead{HR 5183 Value} & 
  \colhead{Unc.} &
  \colhead{HIP 67291 Value} & 
  \colhead{Unc.} &
  \colhead{Unit} &
  \colhead{Unc. Unit}
 }

\startdata
R.A. & 206.74 & 0.034 & 206.87 & 0.042 & deg & mas \\
Dec. & 6.35 & 0.029 & 6.32 & 0.028 & deg & mas \\
Parallax & 31.76 & 0.04 & 31.92 & 0.05 & mas & mas \\
Proper Motion (R.A.) & -510.45 & 0.07 & -509.44 & 0.08 & mas yr$^{-1}$ & mas yr$^{-1}$ \\
Proper Motion (Dec.) & -110.22 & 0.06 & -111.02 & 0.06 & mas yr$^{-1}$ & mas yr$^{-1}$ \\
Radial Velocity & -30.42 & 0.20 & -30.67 & 0.15 & km s$^{-1}$ & km s$^{-1}$ \\
\enddata
\end{deluxetable*}

While the presence of an extremely wide stellar companion to HR 5183 is certainly interesting, HIP 67291 is simply too far away from the planet HR 5183 b to affect its orbit in the current orbital configuration. The median periastron distance of the HIP 67291-HR 5183 orbit (neglecting hyperbolic solutions) is $\sim$10,000 au, well beyond the theorized minimum Sun-Oort cloud separation of 2,000 au \citep{Morbidelli:2005aa}. In the Oort cloud, the galactic potential due to the overall galactic mass distribution is an important driver of orbital evolution, which tells us that even when HIP 67291 is closest to HR 5183 b, its gravitational influence is at most comparable to that of the galactic potential. 

\section{Prospects for Direct Imaging and Detection with Gaia}
\label{sec:astrometry}

Transit probability is given by:

\begin{equation}
    p_{tra} = \left(\frac{R_{*} + R_p}{a} \right)\left( \frac{1+e\sin{\omega}}{1-e^2}\right).
\end{equation} \citep{Winn:2010aa}. Assuming a Jupiter radius for HR 5183 b, $p_{tra}$ = \transitprob{}. Although this probability is lottery-ticket-like, the prospects for detecting HR 5183 b with stellar astrometry and thermal direct imaging are promising. Detection with either of these methods could address the $\sin{i}$ degeneracy, allowing us to obtain a direct mass measurement. 
 
\begin{figure}
    \centering
    \includegraphics[width=0.48\textwidth]{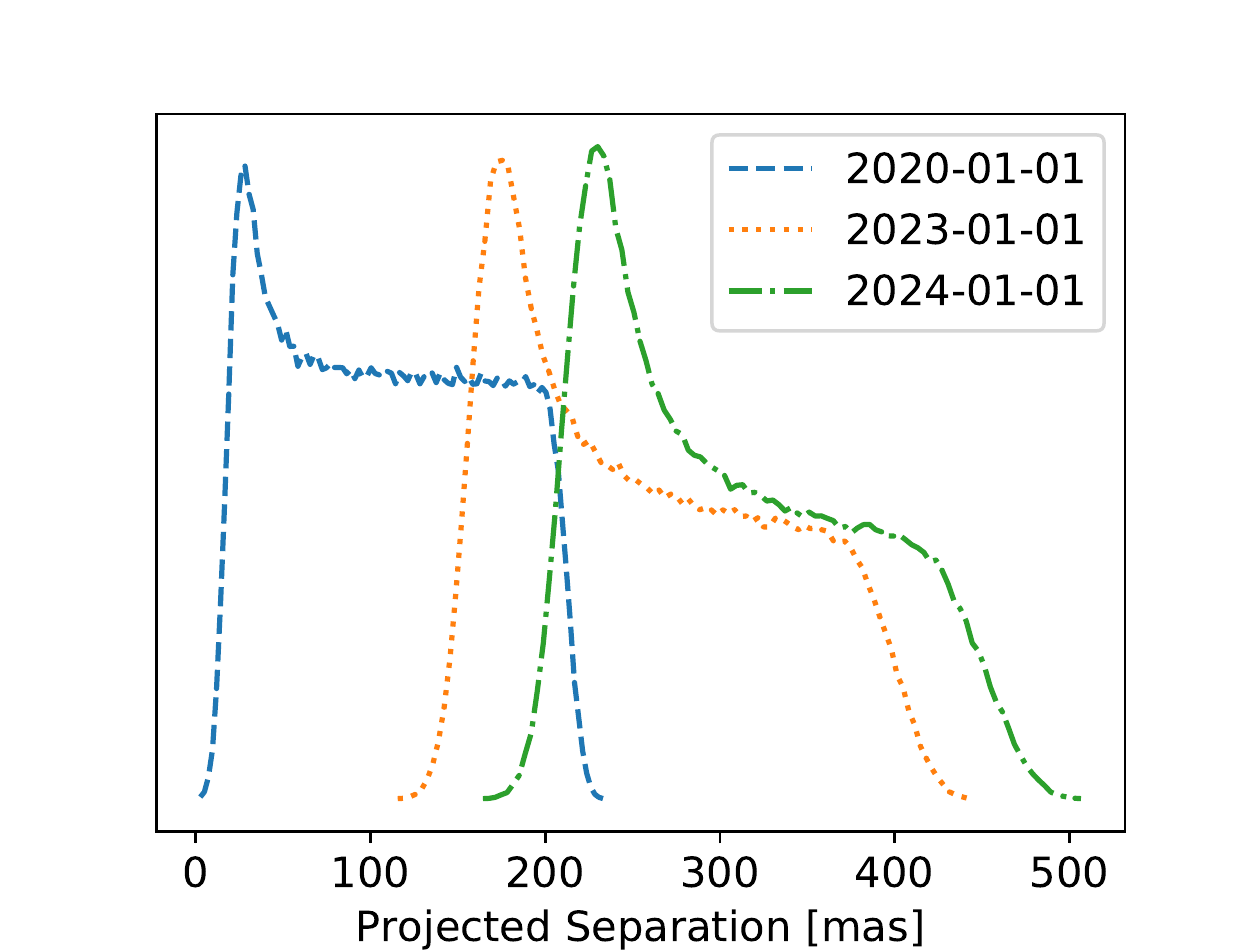}
    \caption{Posterior distributions of the projected separation ($\rho$) of HR 5183 b from its primary at three future epochs. The time of periastron passage is precisely constrained to be 01-2018 from the RVs, and accordingly, the separation posterior generally increases over the next 5 years. In 2024, HR 5183 b will be separated from its host by more than 200 mas with 95\% confidence.}
    \label{fig:sep_pred}
\end{figure}

To investigate prospects for imaging HR 5183 b, we used the orbit-solving code from \texttt{orbitize}\footnote{GitHub.com/sblunt/orbitize} \citep{Blunt:2019aa}, an orbit-fitting toolkit for direct imaging astrometry. First, we determined the angular separation posterior as a function of time. We randomly sampled from the RV orbit posteriors described in the previous section, assigned each sample orbit an inclination (randomly drawn from a uniform distribution in $\cos i$) and an $\Omega$ (randomly drawn from a uniform distribution), and used \texttt{orbitize} to solve for the projected angular separation, $\rho$, at several future epochs. Posterior distributions in $\rho$ calculated using this procedure for three future epochs are shown in Figure \ref{fig:sep_pred}. Since the planet passed periastron so recently, the median of its projected separation posterior generally increases with time over the next 5 years. 

\begin{figure}
    \centering
    \includegraphics[width=0.48\textwidth]{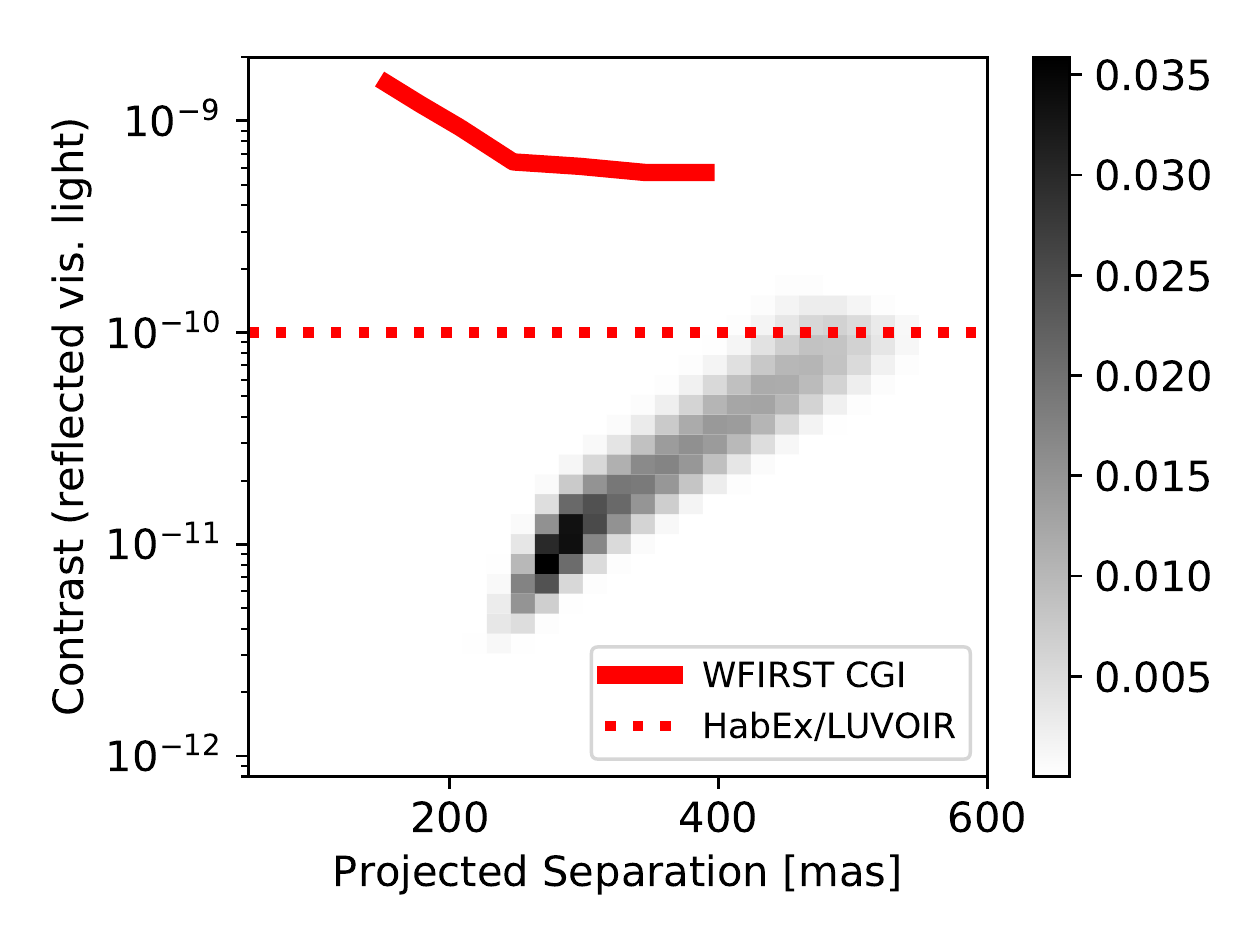}
    \caption{Two-dimensional histogram showing the posterior in planet contrast/projected separation, approximating the planet as a Lambertian disk with an albedo of 0.5. This plot shows a ``snapshot'' of the posterior in time on 01-01-2025. Darker grey indicates higher probability. The predicted post-processing detectability floor for the WFIRST CGI\footnote{Obtained from GitHub.com/nasavbailey/DI-flux-ratio-plot} in narrow FOV mode is shown as a red solid line, where regions above the line are detectable. The dotted line shows the HabEx/LUVOIR performance requirement. HR 5183 b is likely not detectable in reflected light with the WFIRST CGI or HabEx/LUVOIR.}
    \label{fig:vis_contrast_pred}
\end{figure}

\begin{figure}
    \centering
    \includegraphics[width=0.48\textwidth]{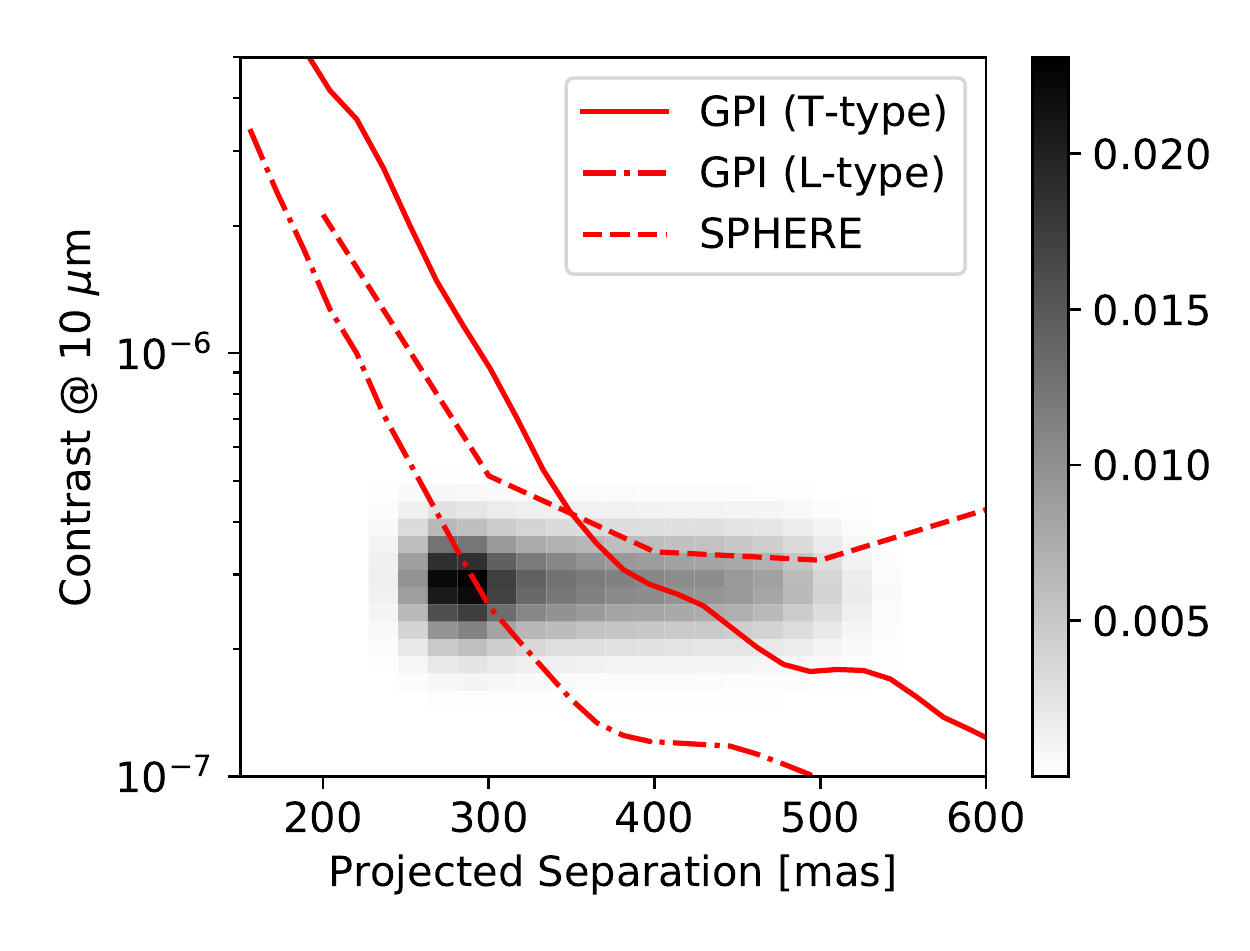}
    \caption{Same as Figure \ref{fig:vis_contrast_pred}, except contrast is calculated in the thermal infrared, assuming $T_{\mathrm{eff}}$ is the equilibrium temperature at periastron. The red lines show the post-processing detectability floors of the Gemini Planet Imager (GPI) and SPHERE, infrared imagers currently in operation (though importantly without 10$\mu$m capabilities). These predictions indicate that HR 5183 b could be imageable with a 10 $\mu$m infrared imager in the next 5 years, bearing in mind the important caveat that the thermal background is higher at 10$\mu$m than at GPI/SPHERE observation wavelengths, potentially negatively impacting contrast capabilities.}
    \label{fig:ir_contrast_pred}
\end{figure}

Next, we calculated contrast posteriors, in both reflected visible and thermal infrared (10 $\mu$m) wavelengths, using the angular separation posterior. To calculate visible reflected-light contrast, we approximated HR 5183 b as a Lambertian disk with an albedo of 0.5, and assumed the star emits as a blackbody. These results are shown in Figure \ref{fig:vis_contrast_pred} on 01-01-2025, along with estimated and required predicted contrast capabilities for future reflected-light coronagraphs. For reference, the phase angle (angle between the observer's line of sight, the planet's location, and the star's location) will be \phaseang{}$\degree$ on this date. Robustly calculating the thermal infrared contrast requires knowing the planet's $T_{\mathrm{eff}}$ (which in turn requires knowledge of non-blackbody effects, such as wavelength-dependent emissivity and age), but as a first optimistic approximation we calculated contrast posteriors assuming the planet emits as a blackbody with temperature given by $T_{\mathrm{eq}}$ at its periastron distance. These results are shown in Figure \ref{fig:ir_contrast_pred}, along with contrast capabilities of two current-generation infrared imagers. In visible reflected light, HR 5183 b appears to be likely beyond the capabilities of even HabEx/LUVOIR, but infrared thermal emission may be a different story. Within 5 years, the planet will most likely be separated from its star by more than 200 mas, and its contrast at 10 $\mu$m, in this optimistic approximation, would be comparable to the performance floors of current-generation infrared imagers like GPI and SPHERE. Instrument concepts like TIKI \citep{Blain:2018aa}, which aim for $1e-7$ contrast at the approximate projected separation of HR 5183 b, are well-suited for this endeavor. 

Another imminent dual-detection prospect for HR 5183 b is with stellar astrometry from Gaia. Gaia will release astrometric timeseries data for HR 5183 with the final data release for the nominal mission (after DR3). To assess potential detectability with Gaia, we similarly randomly sample from the RV orbit posteriors, randomly assign inclinations and $\Omega$ values as described above, and use \texttt{orbitize} to compute relative $\Delta$R.A. and $\Delta$Dec as a function of time for many possible orbits.

\begin{figure*}
    \centering
    \includegraphics[scale=0.75]{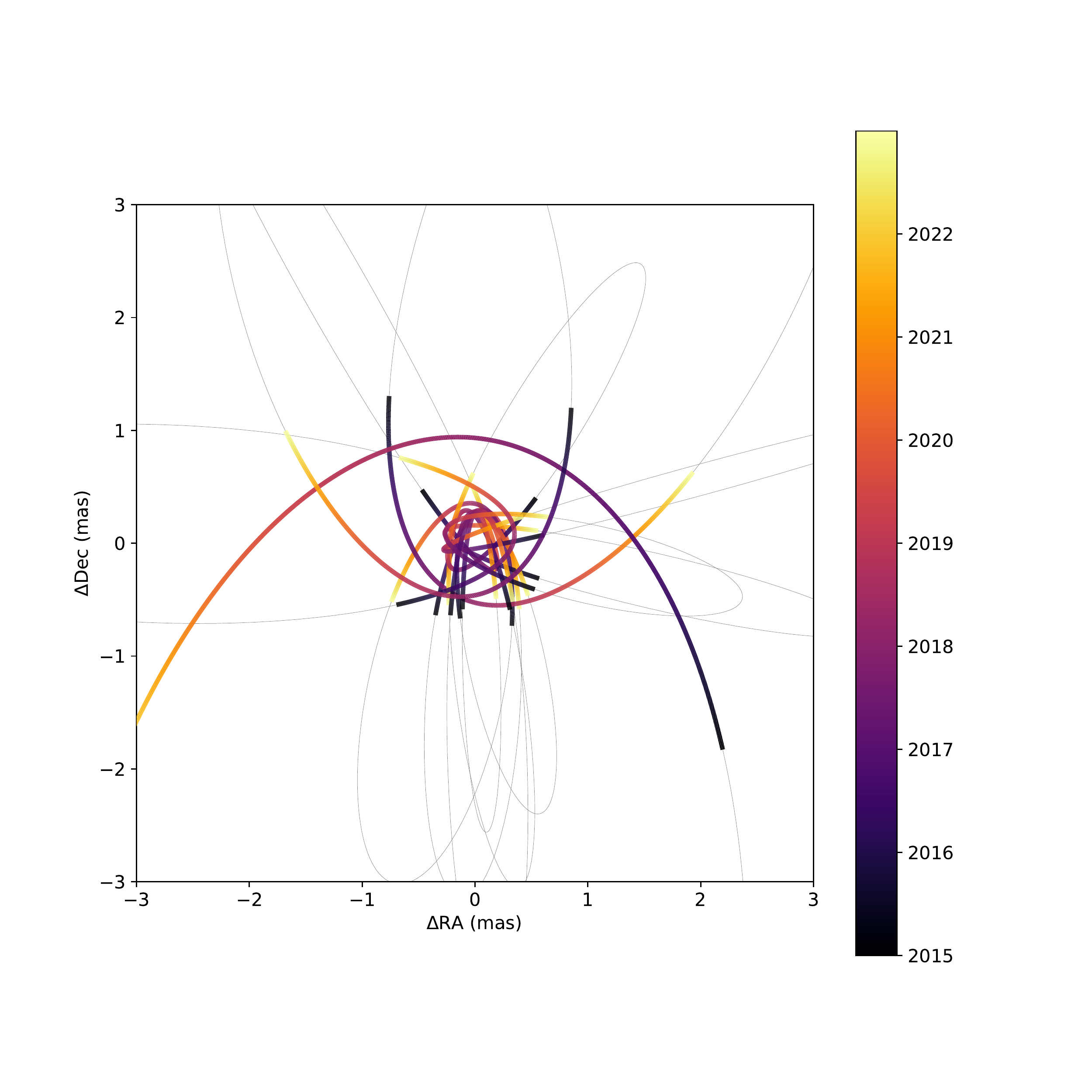}
    \caption{Sky projections of several simulated orbits of the star HR 5183 induced by the planet HR 5183 b. The colored portions of the orbital arcs show elapsed time over the nominal Gaia mission length, with black closest to the present and yellow farthest in the future. The planet's periastron passage in January 2018 is apparent in each orbit.}
    \label{fig:gaia_fireworks}
\end{figure*}

\begin{figure}
    \centering
    \includegraphics[width=0.48\textwidth]{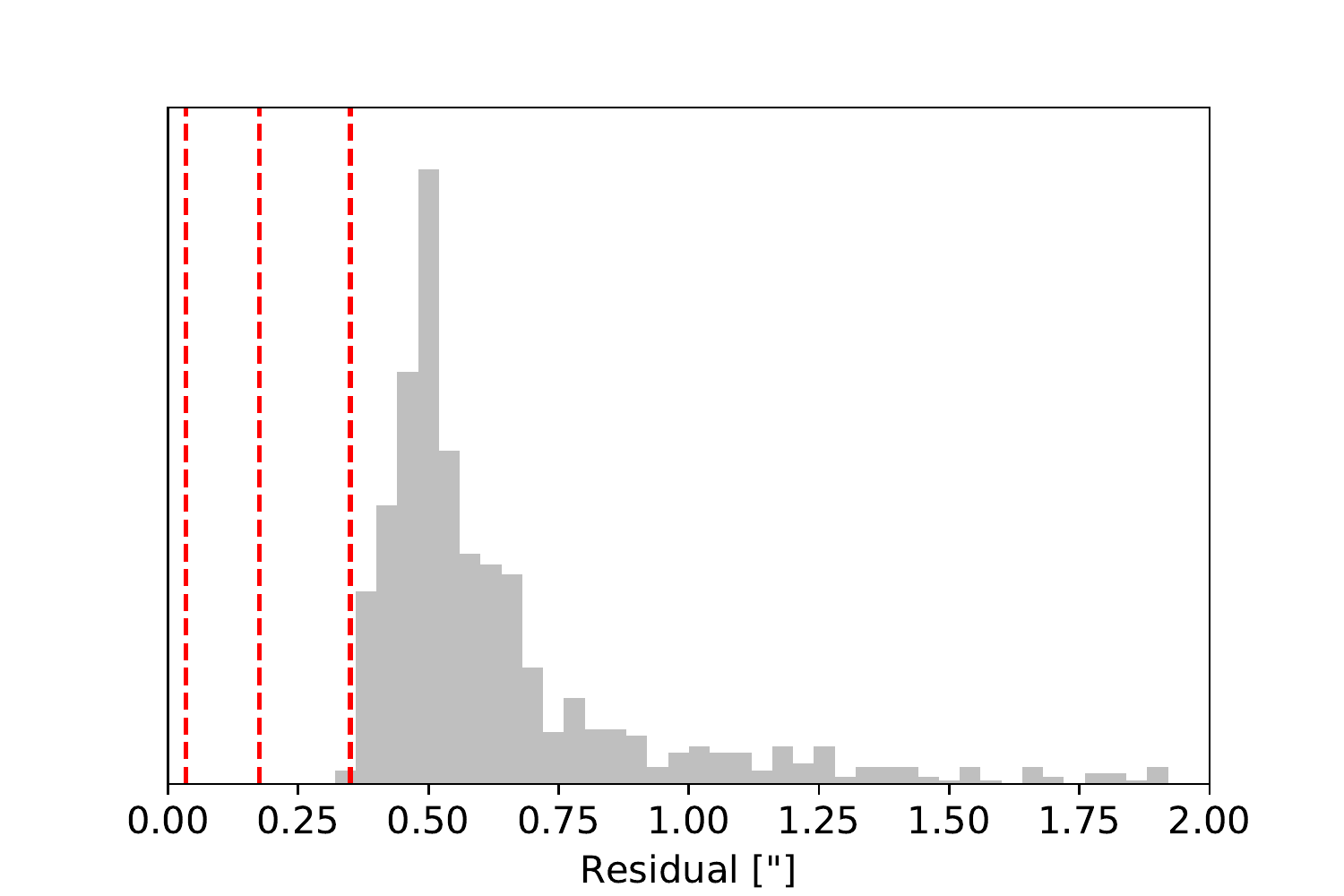}
    \caption{Histogram of the residual values of simulated Gaia astrometric orbits to straight-line fits. To be detectable, the residual must be greater than five times the Gaia measurement uncertainty on the position of HR 5183. Red vertical dotted lines are shown at 1, 5, and 10-$\sigma$. See Section \ref{sec:astrometry} for a complete description of this algorithm. Based on this analysis, 100{}\% of simulated orbits of HR 5183 b are 5-$\sigma$ detectable with Gaia.}
    \label{fig:gaia_resids}
\end{figure}

For HR 5183 b to be detectable with Gaia, its orbit must look sufficiently different from a constant rate of change in $\Delta$R.A. and $\Delta$Dec, which could be interpreted as a proper motion. We therefore fit a line to each generated orbit in our sample (in $\Delta$R.A. and $\Delta$Dec), and subtracted this fit from the sample orbit. If the maximum value of this residual curve exceeded five times the Gaia uncertainty (assumed to be 35 mas, the current astrometric uncertainty for HR 5183 in the Gaia catalogue, and typical for stars of similar spectral type; \citealt{Gaia-Collaboration:2018aa}) in either $\Delta$ R.A. or $\Delta$ Dec, we counted the orbit as detectable (a 5-$\sigma$ detection). We repeated this analysis for 10,000 orbits to estimate the probability of detecting HR 5183 b with Gaia. With this algorithm, we calculate a detection probability of 100\%, or in other words, 100\% of orbits consistent with our RV posteriors will be detectable with Gaia. Representative detectable orbit tracks are plotted in Figure \ref{fig:gaia_fireworks} over the expected Gaia mission length. A histogram of residuals, with the current Gaia uncertainty overplotted, is shown in Figure \ref{fig:gaia_resids}.

Combining the astrometric baselines of Hipparcos and Gaia (using a method similar to \citealt{Dupuy:2019aa}) may also render HR 5183 b detectable in stellar astrometric data, and/or increase the SNR of a Gaia-only detection. Such a project is an excellent avenue for future work on HR 5183 b, especially after the final Gaia nominal mission data release.

\section{Discussion \& Conclusions}
\label{sec:discuss}

HR 5183 b is one of the longest-period exoplanets detected with RVs. Its extreme eccentricity, coupled with one of the longest RV monitoring baselines in exoplanet history, set it apart from other long-period RV-detected planets. Our 22 year observing baseline includes recent observations (2017-2018 season) of the planet's periastron passage, which allowed us to precisely constrain the planet's minimum mass and eccentricity. Without continuous and high-cadence (at least one observation per year) RV monitoring, we would have missed the information-dense periastron passage of HR 5183 b. However, the odds of detecting such an event in a single star with our observing strategy, which has included regular monitoring of $>$100 bright stars for 20 years, are relatively high (roughly 1/3); survey design, rather than serendipity, is at the heart of this discovery.

With an age of several Gyr, HR 5183 b is distinct from the several Myr-old population of long-period planets detected via direct imaging. In addition, its minimum mass of \msini{} \mjup{} means it is likely much less massive than the directly imaged planets, which tend to be closer to 10 $\mjup$ owing to strong selection effects. HR 5183 b also has a much higher eccentricity and longer orbital period than typical RV-detected planets. Table \ref{tab:long-period-planets} and Figure \ref{fig:long-period-planets} compare its orbital properties and those of similar known exoplanets. Although HR 5183 b has an orbital period and mass that make it more similar to directly imaged planets than RV-detected planets, its age and high eccentricity differentiate it from its directly imaged cousins.

The extreme eccentricity and decades-long orbital period of HR 5183 b, coupled with the existence of a widely separated, eccentric stellar companion (Section \ref{sec:stellar_comp}), raise interesting questions about the system's formation. High eccentricity is a signature of past dynamical interactions \citep{Dawson:2014aa}. Moreover, recent dynamical simulations by \citet{Wang:2018aa} revealed that systems hosting multiple young massive planets, presumably near their formation locations, are likely unstable on Gyr or shorter timescales. Therefore, the HR 5183 system might have initially contained multiple massive planets with moderate eccentricities. Planet-planet interactions in such a system could have ejected some planets and transferred angular momentum to the remaining planet(s), pumping their eccentricities. If this is true, the HR 5183 system could be viewed as the ``fate'' of systems like HR 8799. Dynamical work aiming to distinguish between this and other possible formation scenarios (for example, potential past interactions with HIP 67291) would be an excellent avenue for future studies. It will be interesting to learn whether HR 5183 b represents the eventual evolution of multiple giant planet systems like HR 8799, or if it is in a class all its own.

HR 5183 b is poised for dual detection with both thermal infrared high-contrast imaging and stellar astrometry, either of which would break the $\sin{i}$ degeneracy and enable a direct mass measurement. Further work is needed to more convincingly estimate the planet's $T_{\mathrm{eff}}$ as a function of time (taking into account its orbital location and non-blackbody effects), but our optimistic calculations indicate that the planet will be at a favorable projected separation and contrast at 10 $\mu$m within the next 5 years. In addition, our calculations indicate that HR 5183 b will almost certainly be detectable in Gaia data.

The National Academy of Sciences consensus report on Exoplanet Science Strategy states that a key goal of exoplanet science in the next decade is to ``determine the range of planetary system architectures by surveying planets at a variety of orbital separations and searching for patterns in the structures of multiplanet systems'' \citep{National-Academies-of-Sciences:2018aa}. In Figure \ref{fig:long-period-planets}, we plot semimajor axis versus mass for planets listed in the NASA Exoplanet Archive and HR 5183 b. From this figure, it is clear that the discovery of HR 5183 b furthers the goal of characterization at all orbital separations. Our success in discovering and characterizing HR 5183 b demonstrates that RV surveys are capable of detecting exoplanets in a yet unexplored parameter space of eccentric, long-period giant planets. With this discovery, we continue to uncover the astonishing diversity of planetary systems in our the galaxy. 

\newpage

\begin{floattable}
\begin{deluxetable*}{cccccccc}

\rotate
\setstretch{0.6}

\tablecaption{Physical Parameters of HR 5183 b and Close Exoplanet Relatives \label{tab:long-period-planets}}

\tablehead{\colhead{Name} &  \colhead{Semi-major axis} & \colhead{Mass or $m$sin$i$} & \colhead{Eccentricity} & \colhead{Discovery Method} & \colhead{1st Ref.} & \colhead{Ref.} & \colhead{Notes} \\ 
\colhead{} & \colhead{(au)} & \colhead{($M_J$)} & \colhead{} & \colhead{} & \colhead{} & \colhead{} & \colhead{}} 

\startdata
51 Eri b &  $14^{+7}_{-3}$ & $2\pm1$ & $0.21^{+0.06}_{0.40}$ & Imaging & A & B & \\
HD 95086 b & $62^{+20}_{-9}$ & $5\pm2$ & $<0.205$ & Imaging & C & D & \\
HR 8799 b &  $70.8^{+-.19}_{-0.18}$ & $5\pm1$ & $0.018^{+0.018}_{-0.013}$ & Imaging & E & F & \\
HR 8799 c & $43.1^{+1.3}_{-1.4}$ & $7\pm2$ & $0.022^{+0.023}_{-0.017}$ & Imaging & E & F &  \\
HR 8799 d & $26.2^{+0.9}_{-0.7}$ & $7\pm2$ & $0.129^{+0.022}_{-0.025}$ & Imaging & E & F &  \\      
HR 8799 e & $16.2\pm0.5$ & $7\pm2$ & $0.118^{+0.019}_{-0.013}$ & Imaging & G & F &  \\
$\beta$ Pic b & $11.8\pm0.9$ & $13\pm3$ & $0.24\pm0.06$ & Imaging & H & I &  \\
HIP 65426 b  & $120^{+90}_{−40}$ & $8\pm1$ & $<0.43$ & Imaging & J & K & \\
PDS 70 b  & $\sim22$ &  & $10.5\pm3.5$ & Imaging & L & L & \\
LkCa 15 b & $15.7\pm2.1$ & $6\pm1$ & & Imaging & M & M & \\
GJ 676 A c & $6.6\pm0.1$ & $6.8\pm0.1$ &  & Radial Velocity & N & N &  \\
HIP 5158 c &  $7.7\pm1.9$ & $15.04\pm10.55$ & $0.14\pm0.10$ & Radial Velocity & O & O &  \\
HD 30177 c & $9.89\pm1.04$ & $7.6\pm3.1$ & $0.22\pm0.14$ & Radial Velocity & P & P &  \\
47 UMa d &  $11.6^{+2.1}_{-2.9}$ & $1.64^{+0.29}_{-0.48}$ & $0.16^{+0.09}_{-0.16}$ & Radial Velocity & Q & Q &  \\
HIP 70849 b & $20.25\pm15.75$ & $9\pm6$ & $0.715\pm0.245$ & Radial Velocity & R & R &  \\
DP Leo b  & $8.19\pm0.39$ & $6.05^{+0.47}_{-0.58}$ & $0.39\pm0.13$ & Eclipse Timing Variations & S & T & U \\
HR 5183 b & \sma & \msini & \ecc & Radial Velocity & This paper & This paper &  \\
\enddata

\tablerefs{
A--\citet{Macintosh:2015aa}, 
B--\citet{De-Rosa:2015aa},
C--\citet{Rameau:2013aa},
D--\citet{Rameau:2016aa}, 
E--\citet{Marois:2008aa},
F--\citet{Wang:2018aa},
G--\citet{Marois:2010aa},
H--\citet{Lagrange:2009aa},
I--\citet{Dupuy:2019aa},
J--\citet{Cheetham:2019aa},
K--\citet{Chauvin:2017aa},
L--\citet{Keppler:2018aa},
M--\citet{Kraus:2012aa},
N--\citet{Sahlmann:2016aa},
O--\citet{Feroz:2011aa}, 
P--\citet{Wittenmyer:2017aa},
Q--\citet{Gregory:2010aa},
R--\citet{Segransan:2011aa},
S--\citet{Qian:2010aa},
T--\citet{Beuermann:2011aa}
}

\tablecomments{U--This is a binary star}
\tablecomments{The imaged planets shown in this table are lifted from the top section of Table 1 in \citet{Bowler:2016aa}, which compiles planets with median semimajor axis $<100$ au orbiting main-sequence stars. The recent bona-fide discoveries PDS 70 b and HIP 65426 b have been added. Masses for these planets were taken from \citet{Bowler:2016aa} except for these two recent discoveries. All other information was taken from the references in this table. Planets detected by other methods are those on the NASA Exoplanet Archive (accessed 2019-4-14) with orbital periods $>7000$ d.}
\tablecomments{Objects that present as RV trends or incomplete orbital arcs whose orbital posteriors are unconstrained are not included in this table for simplicity. Our intention here is not to tabulate every potentially similar planet to HR 5183 b, but to compile the parameters of a few illustrative cases. We refer interested readers to references in the introduction for additional examples.}

\end{deluxetable*}
\end{floattable}

\onecolumngrid 
\vspace*{30px}
\acknowledgments{
The authors thank those at academic and telescope facilities whose labor maintains spaces for scientific inquiry, particularly those whose communities are excluded from the academic system. The authors would also like to sincerely thank the referee for a thorough and helpful report that greatly improved the quality of this work. 

The authors note that we refer to this planet colloquially as ``planet pi'' because its minimum mass is consistent with $\pi$\mjup{}. 

S.B. would like to thank Vanessa Bailey, Konstantin Batygin, Dave Charbonneau, Robert De Rosa, Dmitry Savransky, Dimitri Mawet, and Dan Fabrycky for helpful conversations. S.B. is supported by the NSF Graduate Research Fellowship, grant No. DGE 1745303. L.M.W. acknowledges support from the Beatrice Watson Parrent Fellowship and the Trottier Family Foundation. M.R.K is supported by the NSF Graduate Research Fellowhsip, grant No. DGE 1339067. A.\ W.\ H.\ acknowledges NSF award 1517655. AV's work was performed [in part] under contract with the California Institute of Technology (Caltech)/Jet Propulsion Laboratory (JPL) funded by NASA through the Sagan Fellowship Program executed by the NASA Exoplanet Science Institute. G.W.H. acknowledges long-term support from NASA, NSF, Tennessee State University, and the State of Tennessee through its Centers of Excellence program.

The McDonald Observatory planet search is supported by the National Science Foundation through grant AST-1313075. The authors thank Ivan Ramirez, Stuart Barnes, and Diane Paulson for for their help with some of the Tull spectrograph observations.

Based, in part, on observations made with the NASA/ESA Hubble Space Telescope, obtained at the Space Telescope Science Institute (STScI), which is operated by the Association of Universities for Research in Astronomy, Inc., under NASA contract NAS 5-26555. These observations are associated with program $\#$'s 12228, 14714 and 15221. Support for these programs was provided by NASA through grants from STScI.

Finally, the authors wish to emphasize the very significant cultural role and reverence that the summit of Maunakea has long had within the indigenous Hawaiian community. We acknowledge that data used in this paper were collected on lands belonging to the K\={a}naka Maoli.

}

\bibliographystyle{aasjournal}
\bibliography{120066}

\appendix

\begin{appendices}
%\twocolumngrid
\section{Photometric Observations \& Transit Search}
\label{sec:photom}

We have been acquiring nightly photometric observations of HR~5183 each year since 2002 with the Tennessee State University (TSU) T8 0.80~m Automatic Photoelectric Telescope (APT) located at Fairborn Observatory in southern Arizona.  The T8 APT is equipped with a two-channel precision photometer that uses a dichroic filter and two EMI 9124QB bi-alkali photomultiplier tubes to separate and simultaneously measure the Str\"omgren $b$ and $y$ pass bands.  The observations were made differentially with respect to three comparison stars, corrected for atmospheric extinction, and transformed to the Str\"omgren photometric system.  The resulting precision of the individual differential magnitudes ranges between $\sim0.0010$ and $\sim0.0015$~mag on good nights, as determined from the nightly scatter in the comparison stars.  Seasonal means of the best comparison stars scatter about their grand means with typical standard deviations of $\sim0.0002$ mag.  Further details on the T8 APT, precision photometer, and the observing and data reduction procedures can be found in \citet{Henry:1999aa}.

HR~5183 has been observed as part of TSU's long-term photometric monitoring program of Sun-like stars \citep{Radick:2018aa}.  These authors report the intrinsic variability in the {\it year-to-year} mean brightness of HR~5183 to be only 0.00028~mag (see their Table~2).  We looked for {\it night-to-night} variability within each of the 17 observing seasons by computing differential magnitudes of HR~5183 versus the mean brightness of the two best comparison stars.  The standard deviations of the nightly observations within each season ranged from 0.00079~mag to 0.00152~mag, with a mean of 0.00106~mag.  These values are consistent with the nightly precision of our observations.  We also computed power spectra for each observing season and found no evidence for any periodicity between 1 and 200 days.  We conclude that HR~5183 is constant to high precision on both nightly and yearly timescales. 

Finally, we searched the complete 17-year data set for possible transits of unknown planets with orbital periods between 1.5 and 200~d, using a simple box-fit, matching-filter technique.  No evidence for any transits, to a limit of $\sim0.002$~mag was detected in our photometry.

\section{Details about Coronagraphic Images}
\label{sec:images}

We analyzed two sets of images of HR 5183 to search for additional stellar and substellar companions in the system, ultimately finding no evidence for additional companions. In the sections below, we describe the reduction of these images and calculation of contrast curves to describe our sensitivity.

\subsection{HST Images}

HR 5183 was observed with HST/STIS in GO programs 12228, 14714, and 15221 (G. Schneider, PI). For GO program 12228, HR 5183 was observed in 2011 as a color-matched PSF template star for the reduction and imaging of the circumstellar disk of HD 107146. HR 5183 was observed identically at two different field orientation angles (approximately 2 months apart, contemporaneous with the HD 107146 observations). At each epoch, two STIS occulting apertures, WedgeA-0.6, and WedgeA-1.0, were  used sequentially over a single spacecraft orbit. These observations of HR 5183 were not planned for rotational differential imaging (RDI) reduction and analysis, and are non-optimal for that purpose. Our interest in detecting and characterizing any companions to HR 5183 nonetheless motivates our analysis of these archival images. 
% to account for the large angular size of the HD 107146 disk to which the template PSFs were later applied. 
% The WedgeA-0.6 occulter provides a smaller inner working angle, while WedgeA-1.0 is useful for probing further from the occulted star with greater depth of integration.
The observations are listed in Table \ref{tab:hstdata}; for more details see \citet{Schneider:2014aa}.
%To better stabilize the PSF for intended 2RDI, the two orbits would have been planned uninterrupted (except an inter-orbit Earth occultation) in "back-to-back" visibility periods, with no spacecraft repointing (other than roll) between, and not separated in time.

Basic image reduction was performed as described in \citet{Schneider:2014aa}. In brief, all raw images were calibrated using the Space Telescope Science Data Analysis System (STSDAS) \texttt{calstisa} software\footnote{http://www.stsci.edu/hst/stis/software/analyzing}, and the same-orbit/same-occulter images were median combined into four single ``visit-level'' images. Same-occulter image pairs from the different visits were astrometrically aligned to minimize PSF-subtraction residuals using the IDP3 image analysis package\footnote{https://archive.stsci.edu/prepds/laplace/idp3.html} \citep{Stobie:2006aa}. WedgeA-0.6 to WedgeA-1.0 image alignments were established using the ``X-marks the spot'' diffraction spike centroid method \citep{Schneider:2014aa} for the unsubtracted images.

We define the image contrast at any stellocentric angular distance (r) to be the ratio of the flux density contained within any pixel in the unocculted stellocentric field to that of the flux density in the (occulted) central pixel in the stellar PSF. As the latter is not directly measurable from these data, we used the high-fidelity TinyTim telescope and instrument PSF modeling code \citep{Krist:2011aa} as codified in the STIS imaging Exposure Time Calculator\footnote{http://etc.stsci.edu/etc/input/stis/imaging} to calculate the image contrast. This analysis informs us that 22\% of the V=6.3, G0V, stellar flux produced by HR 5183 would be contained in the central 50.\arcsec077 square pixel of the (occulted) PSF. % and, for the instrumental configuration employed, would produce an instrumental count rate in that pixel of 4.268E6 counts second-1 pixel-1 (with GAIN = 4, used for all observations). 
We compared this to the 1-$\sigma$ standard deviations in the subtraction-nulled 2RDI signal at incrementally increasing stellocentric annuli of 1-pixel widths to produce a contrast curve (1-$\sigma$ point-source detection limit in contrast versus. stellocentric angle). The results for both wedges used are shown in Figure \ref{fig:images}.

In HST Cycles 24 and 25, HST GO programs 14714 and 15221 (G. Schneider, PI) episodically monitored the HD 107146 debris disk with STIS PSF-template subtracted coronagraphy. These programs collected four additional epochs of coronagraphic observations of HR 5183 (see Table \ref{tab:hstdata}) to serve as contemporaneous PSF subtraction templates. These images were analyzed as described above. % the resultant contrast curve for all of the observation dates are shown in~\autoref{fig:hstall}.
%These observations that differed from the original GO 12228 observations with: (a) only with WedgeA1.0, also by need and design, saturating the inner regions of those PSFs, (b) with longer intra-observational cadence of 4 - 8 months, were also not designed (otherwise properly with "back-to-back" orbits) for 2RDI but, were evaluated here as a serendipitous opportunity.
These observations had longer intra-observational cadence (4-8 months) and did not improve the contrast curve from GO 12228. 

\subsection{NaCo data}

HR 5183 was observed with NaCo in Ks-band on 2007 July 2 \citep{Lenzen:2003aa} as a PSF-reference star for VLT program 079.C-0420 (M. Clampin, PI). Observations used the 1$\farcs$4 diameter coronagraphic mask and S27 camera, whose 1024x1024 pixel size provided a 28$\farcs$ x 28$\farcs$ field of view given the pixel scale of 27.15 mas pixel$^{-1}$. The observing sequence included three science frames and three sky frames, each with exposure times of 30 seconds. We also used a master twilight flat and a bad pixel array. 
    
We median combined the sky images into a single master sky frame and subtracted this from each science frame, then applied a flat-field correction using the master flat. We used the bad pixel array to identify erroneous pixel counts, and replaced these with the median values of adjacent pixels. One of the three science frames was taken with a position angle of 30\degree with respect to the other two images, which necessitated registration of the images and de-rotation of the frame. We used the non-linear least squares optimization routine in SciPy\footnote{https://docs.scipy.org/doc/scipy/reference/optimize.html} to fit a two-dimensional Moffat function \citep{Moffat:1969aa} to the wings of the stellar PSF extending beyond the coronagraphic mask. Each image was aligned by the determined stellar position and de-rotated based on the relative position angle. The three aligned science frames were then median combined to produce a single composite frame, which was subsequently cropped to 800x800 pixels.

The flux of the stellar PSF and the photometric noise were approximated by dividing the image into concentric annuli about the determined stellar center and computing the median and standard deviation of the counts within each annulus. Regions with contributions from the bright diffraction spikes in the images were excluded from this calculation. Additionally, prior to calculation of each standard deviation, a $\pm10\sigma$ sigma-clipping routine was applied to each set of counts to eliminate cosmic rays and bad pixels. To create a stellar PSF model and noise model, the sets of medians and standard deviations were interpolated over the angular separation of each pixel from the target star. The final image was then created by subtracting the stellar PSF model from the composite science image. The result contains flux from diffraction spikes and some residual speckle noise near the coronagraphic mask, but no obvious point source companions.
    
To characterize the detection limit of these data, we computed a contrast curve following \citealt{Uyama:2017aa}. We adopt the computed 1-$\sigma$ noise at each position as the point-source detection limit. We estimated the magnitude of the unobstructed target star using the 2MASS Ks magnitude of HR 5183 and the zero-point of nightly NaCo standard star observations \citep{Skrutskie:2006aa}. The total counts from each threshold artificial source were approximated as the counts from a two dimensional Gaussian with amplitude equal to the computed threshold amplitude and width equal to the width reported for nightly standard star observations. This was then divided by the exposure time to achieve counts per second for the detection threshold at each position. The detectable contrast at each separation was taken to be the ratio of the target count rate to the count rate of each threshold source. These results are shown in Figure \ref{fig:images}.

\begin{deluxetable}{lrrcc}
\tablecaption{HST Observations of HR 5183 \label{tab:hstdata}}
\tablehead{
  \colhead{Visit Id} &   \colhead{Date (UTC)} &   \colhead{Occulter} &   \colhead{Exptime (s)} &   \colhead{Orientat.} }
\startdata
OBIW23 & 2011-05-03 & WedgeA-0.6 & 24 x 11s & 85.978\degree\\
OBIW23 & 2011-05-03 & WedgeA-1.0 &   7 x 206.5s & 85.978\degree\\
OBIW27 & 2011-02-22 & WedgeA-0.6 & 24 x 11s & 233.038\degree\\
OBIW27 & 2011-02-22 & WedgeA-1.0 &   7 x 206.5s & 233.038\degree\\
OD5403 &  2017-03-26 &  WedgeA-1.0 &  10 x 206.5 &    204.093\degree\\
OD5413 &  2017-07-28 &  WedgeA-1.0 &  10 x 206.5 &  64.6028\degree\\
ODH403 &  2018-03-16 &  WedgeA-1.0 &  10 x 206.5 &  217.580\degree\\
ODH413 &  2011-02-22 &  WedgeA-1.0 &  10 x 206.5 &  64.709\degree\\
\enddata
\end{deluxetable}

\end{appendices}

\end{document}